\documentclass{article}

\usepackage[preprint, nonatbib]{neurips_2023}
\usepackage{graphicx}
\usepackage{subcaption}

\usepackage{setspace}
\usepackage{multirow}
\usepackage{verbatim}

\usepackage[utf8]{inputenc} 
\usepackage[T1]{fontenc}    
\usepackage{hyperref}       
\usepackage{url}            
\usepackage{booktabs}       
\usepackage{amsfonts}       
\usepackage{amsmath}       
\usepackage{nicefrac}       
\usepackage{microtype}      
\usepackage{xcolor}         

\title{Physics-Informed Machine Learning Approach in Augmenting RANS Models Using DNS Data and DeepInsight Method on FDA Nozzle}

\author{%
Hossein Geshani \quad
Mehrdad Raisee Dehkordi \quad
Masoud Shariat Panahi \\
School of Mechanical Engineering, University of Tehran, Tehran, Iran \\
\texttt{hossein.keshani@ut.ac.ir} \\
}

\begin{document}

\maketitle

\begin{abstract}
We present a data-driven framework for turbulence modeling, applied to flow prediction in the FDA nozzle. 
In this study, the standard RANS equations have been modified using an implicit-explicit hybrid approach. New variables were introduced, and a solver was developed within the OpenFOAM framework, integrating a machine learning module to estimate these variables. The invariant input features were derived based on Hilbert's basis theorem, and the outputs of the machine learning model were obtained through eigenvalue-vector decomposition of the Reynolds stress tensor.
Validation was performed using DNS data for turbulent flow in a square channel at various Reynolds numbers. A baseline MLP was first trained at $Re=2900$ and tested at $Re=3500$ to assess its ability to reproduce turbulence anisotropy and secondary flows. To further enhance generalization, three benchmark DNS datasets were transformed into images via the Deep-Insight method, enabling the use of convolutional neural networks. The trained Deep-Insight network demonstrated improved prediction of turbulence structures in the FDA blood nozzle, highlighting the promise of data-driven augmentation in turbulence modeling.
\end{abstract}

\section{Introduction}

\begin{figure}[h!]
\centering
\includegraphics[width=\textwidth]{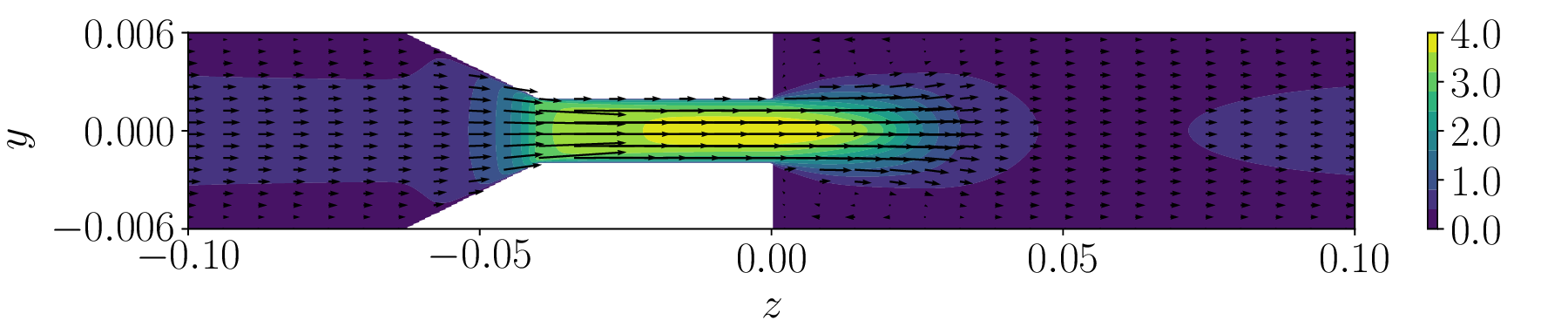}
\caption {Velocity field obtained from RANS simulation for blood measurement nozzle}
\label{fig:FDA_RANS_Velocity}
\end{figure}

In 2009, the FDA hosted an interlaboratory study to evaluate computational fluid dynamics (CFD) methods in the safety investigation of healthcare equipment. The benchmark study focused on a general healthcare device featuring a cylindrical nozzle with a diameter of 0.012 m, characterized by a sudden contraction and a conical diffuser of 10 degrees on both sides of a throat with a length of 0.04 m and a diameter of 0.004 m(Fig. \ref{fig:FDA_RANS_Velocity}). This study is pivotal because accurate flow predictions are essential for assessing blood damage criteria, such as hemolysis, which can directly impact patient safety and treatment efficacy. Various numerical simulations were conducted, and their outcomes were compared with planar particle imaging velocimetry (PIV) data across laminar, transitional, and turbulent flow regimes  \cite{10.1115/1.4003440,10.1007/s13239-012-0087-5}. Notably, the CFD results, particularly in turbulent regimes, did not align well with the PIV measurements, highlighting the need for enhanced predictive capabilities. In some cases, CFD predictions of the maximum wall shear stress in sudden contraction showed significant discrepancies compared to experimental results, which is considered as a major source term for the hemolysis criteria in healthcare devices. 

Simulating biological effects like hemolysis, which involves red blood cell damage in medical devices, requires highly accurate predictions of flow and stress fields. These fields are essential for accurately solving equations related to hemolysis. Despite some limitations, Reynolds-Averaged Navier-Stokes (RANS) turbulence models are widely used in industry because they strike a balance between accuracy and computational efficiency. However, RANS models can have significant errors in predicting key factors like Reynolds stresses, turbulent viscosity, and velocity fields, depending on the specific case. These errors are often due to assumptions made in Boussinesq's hypothesis and constants used in simple one- or two-equation models, which are derived from calibrations based on standard fluid mechanics problems. By improving the accuracy of Reynolds stress and velocity predictions, we can expect more accurate estimates for hemolysis and other blood damage criteria.

In fluid dynamics, various types of data are utilized, including laboratory results, field measurements, and numerical simulations. The emergence of Big Data has significantly influenced fluid mechanics in recent years, driven by advancements in high-performance computing and enhanced experimental techniques \cite{10.1007/978-3-319-41217-7}. Machine learning algorithms, encompassing supervised, weak supervision/semi-supervised, and unsupervised learning, are increasingly applied in fluid mechanics, providing a versatile modeling framework that addresses various challenges, including the turbulence modeling closure problem. 
The discrepancies observed between RANS predictions and PIV measurements in the FDA benchmark nozzle illustrate the limitations of traditional turbulence models in accurately capturing complex flow dynamics in medical devices. This discrepancy underscores the importance of developing turbulence models, such as Reynolds-Averaged Navier-Stokes (RANS), that can better predict flow characteristics, Reynolds stress, and shear stress using machine learning (ML) approaches that leverage direct numerical simulation (DNS) data from benchmark problems.

Researchers have utilized both offline data (DNS datasets from unrelated flows) and online data (real-time data from the target flow) to enhance predictions in fluid dynamics \cite{Dow2011QuantificationOS}, \cite{10.1016/j.jcp.2016.07.038} \cite{Iungo_2015}. For instance, in the study by Dow \cite{Dow2011QuantificationOS}, DNS data of straight channel flows was utilized to identify turbulence viscosity differences, represented as $\Delta\nu_{t}$, modeled with the $k-\omega$ approach. This pioneering work sets the stage for future research by demonstrating how advanced datasets can lead to refined turbulence models. To extend this model to channels with wavy walls, researchers represented the logarithm of the viscosity differences as Gaussian random fields and averaged these over the domain to capture flow characteristics generically.
In a related approach, Duraisamy et al. \cite{Parish2016APF}, \cite{10.1063/1.4947045} introduced the $\beta$ parameter as a location-specific correction factor in the source terms of the transport equations. In their model, $\beta$ serves as a multiplicative factor in the production terms for $\widetilde{\nu}_t$ in the Spalart-Allmaras model and for $\omega$ in the $k-\omega$ model, enabling regional adjustments that improve alignment with DNS data. This innovative approach highlights the potential for localized modifications to enhance turbulence modeling accuracy.
By calibrating the $\beta$ parameter, they were able to assess and reduce uncertainties in the turbulence models, advancing RANS predictions for similar flows. Xiao et al. \cite{10.1016/j.jcp.2016.07.038} further extended this work by using real-time velocity dispersion measurements to quantify discrepancies in the Reynolds stress tensor, denoted as $\Delta \tau_{\alpha}$, predicted by RANS models. Their method allowed for the estimation of additional physical quantities, such as turbulence kinetic energy and Reynolds stress anisotropy, through sparse velocity measurements, adding further depth to RANS model correction efforts.

The methods discussed provide a foundational approach for achieving a central objective: predicting turbulence with standard RANS models using well-validated offline data, such as direct numerical simulation (DNS) or laboratory measurements. By calibrating the difference between true Reynolds stress and that derived from RANS models, these methods aim to extend accurately to similar flow scenarios. Wu et al. \cite{Wu2015ABC}, following the approach by Xiao et al. \cite{10.1016/j.jcp.2016.07.038}, showed that calibrating the difference in Reynolds stress based on limited velocity data could be applied to flows with Reynolds numbers much higher than the initial reference cases. This method significantly improved the accuracy of velocity and other flow predictions, underscoring the promise of using data-driven models for turbulence prediction. However, the approach taken by Wu et al. \cite{Wu2015ABC} had an inherent limitation: it assumes that $\Delta \mathbf{\tau}_{\alpha}$, the discrepancy in Reynolds stress, depends solely on physical coordinates, $\mathbf{x}$, meaning the discrepancy calibration holds only for flows within the same geometric configuration at corresponding locations. Consequently, attempts to generalize across geometries—such as from square to rectangular cross-sections—were less successful. This constraint is shared by Dow and Wang’s approach \cite{Dow2011QuantificationOS}, where the Gaussian random fields they propose also directly depend on spatial coordinates.
Although Wu's method has certain limitations, its effectiveness largely stems from defining key variables—like anisotropy and directional parameters—using the eigenvalues and eigenvectors of the Reynolds stress matrix, rather than the matrix elements directly. To overcome the limitation in Wu's calibration-prediction framework, an improved approach would construct the discrepancy functions in terms of selected feature variables $\mathbf{q}$ rather than relying on physical coordinates $\mathbf{x}$. By defining $\Delta \mathbf{\tau}_{\alpha}$ as a function of $\mathbf{q}$, this approach aims to retain the strengths of Wu’s method while broadening the applicability of calibrated differences across a wider range of flow scenarios.

In a similar direction, Duraisamy et al. \cite{Duraisamy2015NewAI} utilized dimensionless flow properties for their input space, which avoids direct reliance on physical coordinates. However, their input features lacked Galilean invariance. Ling et al. \cite{10.1016/j.jcp.2016.05.003} addressed this by demonstrating that machine learning models perform more accurately with Galilean-invariant input features. Building on this concept, Ling and Templeton \cite{10.1063/1.4927765} proposed a set of 12 Galilean-invariant features for use with a random forest classifier, while Wang et al. \cite{Wang_2017} applied these features in a physics-informed machine learning (PIML) framework. This approach allowed for learning Reynolds stress discrepancies within the mean flow’s feature space, making it possible to predict Reynolds stresses in new flows without additional data.
Wang and colleagues’ research evaluated PIML’s performance specifically in predicting Reynolds stress tensors but did not explore its impact on velocity fields. They noted several challenges, particularly in incorporating modified Reynolds stresses into a RANS solver to accurately predict mean velocity and pressure fields. The study underscored the importance of high-quality training data to produce reliable velocity field predictions. Additionally, achieving high quality Reynolds stress predictions necessitates accurate point predictions and high-quality Reynolds stress derivatives, due to the influence of Reynolds stress gradients on velocity and pressure fields. Purosawa and Thompson \cite{10.1063/1.4966639, 10.1016/j.compfluid.2016.01.014} found that explicitly integrating the Reynolds stress tensor into numerical models can cause issues with convergence. This is because directly replacing the Reynolds stress tensor in equations leads to rapidly compounding errors, resulting in instability. To address this, they suggest incorporating implicit terms, which can help stabilize the solution when working with Reynolds stress tensors.

This paper adopts the approach of Xiao, validating against channel flow with a square cross-section while aiming to enhance RANS turbulence model predictions through the integration of machine learning (ML) approaches. The research specifically focuses on employing supervised algorithms to achieve more accurate estimates of the Reynolds stress tensor in RANS modeling, utilizing a more extensive direct numerical simulation (DNS) dataset from similar benchmark problems. The goals include developing a modified RANS-based solver that integrates ML to adjust model predictions, validating this solver using established DNS data, and training a more advanced ML model, such as DeepInsight, using a validated DNS database for application to more complex flows like the FDA nozzle, which comprises three sections of simpler flows. The modified solver is built using OpenFOAM’s object-oriented framework, while the ML algorithms are implemented with TensorFlow. The PythonCAPI interface is employed to link the solver with the ML model, facilitating the transfer of both implicit and explicit corrector terms related to the Reynolds stress tensor.

\section{Methodology}
This section aims to investigate the process of reaching the modified RANS equations so that the effect of non-linear terms not seen in the linear eddy-viscosity assumption can be included in the Reynolds stress tensor.
Starting from the Navier-Stokes and continuity equations, and using the Reynolds averaging method, we will reach RANS equations:
\begin{equation}\label{RANS}
\frac{\partial}{\partial t}(\rho \overline{\mathbf{u}})+\nabla \cdot(\rho \overline{\mathbf{u}} \otimes \overline{\mathbf{u}})=\mathbf{g}+\nabla \cdot (\overline{\tau})-\nabla \cdot(\mathbf{\tau_R})
\end{equation}
Where $\overline{\mathbf{u}}$ is the mean time-averaged velocity vector, $\overline{\tau}$ is the molecular averaged stress tensor of a Newtonian fluid, and 
$\mathbf{\tau_R}=-\rho \overline{v_{i}^{'} v_{j}^{'}}$
represents the Reynolds Stress Tensor (RST).

The following relation determines a Newtonian fluid's molecular averaged stress tensor.
\begin{equation}
\bar{\tau}=-\left(p+\frac{2}{3} \mu \nabla \cdot \overline{\mathbf{u}}\right) \mathbf{I}+\mu\left(\nabla \overline{\mathbf{u}}+(\nabla \overline{\mathbf{u}})^T\right)
\end{equation}

The general form of algebraic eddy viscosity model for turbulent stresses can be expressed as \cite{pope_1975}:
\begin{equation}\label{eq:algRST}
\begin{aligned}
\mathbf{b}(\mathbf{S}, \boldsymbol{\Omega})=& \sum_{n=1}^{\mathbf{10}} G^{(n)} \mathcal{T}^{(n)} \\
=& G^{(1)} \mathbf{S}+G^{(2)}(\mathbf{S} \boldsymbol{\Omega}-\mathbf{\Omega} \mathbf{S})+G^{(3)}\left(\mathbf{S}^2-\frac{1}{3} \operatorname{tr}\left(\mathbf{S}^2\right) \mathbf{I}\right) \\
&+G^{(4)}\left(\mathbf{\Omega}^2-\frac{1}{3} \operatorname{tr}\left(\boldsymbol{\Omega}^2\right) \mathbf{I}\right)+\cdots
\end{aligned}
\end{equation}
which $\mathbf{b}$ is the deviatoric Reynolds stress tensor , and 
$\mathbf{S}$ 
and 
$\mathbf{\Omega}$
are rotation-rate and strain-rate tensors respectively.
\begin{equation}\label{eq:dev_RST}
\mathbf{b} = \frac{1}{2 \rho k}\left(\mathbf{\tau_R}-\frac{2}{3} \rho k \mathbf{I}\right)
\end{equation}
\begin{equation}
\mathbf{S} = \frac{1}{2} (\frac{\partial\overline{v_i}}{\partial x_j} + \frac{\partial\overline{v_j}}{\partial x_i})
\end{equation}
\begin{equation}
\mathbf{\Omega} = \frac{1}{2} (\frac{\partial\overline{v_i}}{\partial x_j} - \frac{\partial\overline{v_j}}{\partial x_i})
\end{equation}

Recently, researchers have observed that solving the RANS equations with explicit Reynolds stresses can amplify small errors in the Reynolds stresses into significant errors in the mean velocity. Thompson et al. extended this analysis to flow in a square cross-section channel, substituting Reynolds stresses from several well-known DNS databases across a wide range of frictional Reynolds numbers, and observed similar convergence issues. In order to overcome convergence challenges due to explicit treatment of RST one can introduce implicit terms starting from Eq. \ref{eq:dev_RST} \cite{Wu_2018}:
\[\mathbf{b} = -\nu_{t}^{L} \mathbf{S} + \nu_{t}^{L} \mathbf{S} + \mathbf{b}\]
\[ \mathbf{b} = -\nu_{t}^{L} \mathbf{S} + \nu_{t}^{L} \mathbf{S} +  \frac{1}{2\rho k}\left(\mathbf{\tau_R}-\frac{2}{3}\rho k \mathbf{I}\right)\]
\begin{equation}\label{eq:RST_dev_mod}
\mathbf{b} = \nu_{t}^{L} \mathbf{S} + [\mathbf{\tau_{R}}/(2\rho k) - \mathbf{I}/3 - \nu_{t}^{L} \mathbf{S}]
\end{equation}
we simplify Eq. (7) by defining a linear Reynolds stress tensor that aligns with the original definition of Reynolds stress:
\begin{equation}\label{tau_L_def}
\mathbf{\tau_{L}} = -2\rho \nu_{t}\mathbf{S} + \frac{2}{3}\rho k\mathbf{I}
\end{equation}
Based on the initial definition of the deviatoric Reynolds stress tensor, the following relation is obtained for the modeled Reynolds stress tensor:
\begin{equation}\label{RST_modeled}
\mathbf{\tau_{m}} = -2\rho \nu_{t}\mathbf{S} + [\mathbf{\tau_{R}} - \mathbf{\tau_{L}}] + \frac{2}{3}\rho k\mathbf{I}
\end{equation}
Therefore, with the symbolic definition of the Reynolds stress difference obtained from RANS simulation and the true DNS Reynolds stress, a relation for the modified Reynolds stress can be obtained.
\begin{equation}\label{RST_modeled_mod}
\mathbf{\tau_{m}} = -2\rho \nu_{t}\mathbf{S} + [(\tau^{RANS} + \Delta\mathbf{\tau}) - (\mathbf{\tau}^{RANS} + \Delta\mathbf{\tau}^L)] + 2/3\rho k_m \mathbf{I}
\end{equation}

A simple mapping function is considered based on the input variables mentioned in the Eq. \ref{eq:algRST}, and to make it more generalized, one could impose the effects of pressure gradient and turbulent kinetic energy too \cite{10.1103/physrevfluids.2.034603}.
\begin{equation}
\mathbf{\tau_{m}} = g(\mathbf{S},\mathbf{\Omega}, \nabla p, \nabla k)
\end{equation}
where 
$\mathbf{Q} = \{\mathbf{S},\mathbf{\Omega}, \nabla p, \nabla k\}$
is comprised of input features.
The Ling normalization method \cite{10.1063/1.4927765} is adopted to ensure the nondimensionality of the inputs. 

As in traditional turbulence modeling, it is desirable that the form of the Reynolds stress mapping function should be invariant to changes in the coordinate system, this property should also be maintained in data-based turbulence modeling. Therefore, the form of the function
$\mathbf{\tau_{m}}= g(\mathbf{S}, \mathbf{\Omega}, \nabla p, \nabla k)$
must be invariant under rotational and reflective transformations of the coordinate system or Galilean transformation of the reference coordinate system.

To ensure the rotational invariancy of the $g$ function
$\mathbf{\tau_{m}} = g(\mathbf{S}, \mathbf{\Omega}, \nabla p, \nabla k)$
under arbitrary rotations of the coordinate system, the following relationship must hold:
\begin{equation}
\mathbf{Q} \mathbf{\tau_{m}} \mathbf{Q}^T
=
g\Big(
  \mathbf{Q S Q}^T,
  \mathbf{Q} \boldsymbol{\Omega} \mathbf{Q}^T,
  \mathbf{Q} \nabla p \mathbf{Q}^T,
  \mathbf{Q} \nabla k \mathbf{Q}^T
\Big)
\end{equation}
where rotation matrix $\mathbf{Q}$ is an orthogonal matrix with determinant 1 (i.e. $\mathbf{Q}^T = \mathbf{Q}^{-1}$). Rotational invariancy of the trained function $g$ is guaranteed by ensuring rotational-invariant inputs and outputs.
Hilbert's basis theorem \cite{enwiki:1158757854} states that for a tensor set with a finite number of members, there exists a set with a finite number of members that are invariant to the rotation mapping \cite{Spencer1962IsotropicIB}. Specifically for the collection
$\{\mathbf{S}, \mathbf{\Omega}, \nabla p, \nabla k\}$
, the integration basis include the traces of all independent matrices that are formed according to the Cayley-Hamilton theorem \cite{enwiki:1157519575}. Applying Hamilton's theorem to this particular set will result in 47 variables.

We also decompose Reynolds stress difference tensors to rotational invariant variables as the model outputs. The eigenvalue-vector decomposition (EVD) has been applied to the deviatoric part of the Reynolds stress due to the rotational invariant property of EVD:
\begin{equation}\label{RST_eigen_decomposition}
\mathbf{\tau} = 2k(\frac{1}{3}\mathbf{I}+\mathbf{b}) = 2k(\frac{1}{3}\mathbf{I} + \mathbf{V }\Lambda\mathbf{V}^{T})
\end{equation}

Where
$\mathbf{V} = [\mathbf{v_1},\mathbf{v_2},\mathbf{v_3}]$
is a matrix whose columns are eigenvectors of $\mathbf{b}$ and
$\Lambda=diag[\lambda_1,\lambda_2,\lambda_3]$
is a matrix whose main diameter are the eigenvalues in which
$\lambda_1+\lambda_2+\lambda_3=0$. Due to the symmetry of the Reynolds stress deviation matrix, the eigenvectors are orthogonal to each other. (The $\mathbf{V}$ matrix is orthonormal.)
According to the property mentioned for the eigenvalues, these values can be mapped to the 2D barycentric coordinate system. This mapping aims to ensure the invariancy of the output variables of the mapping function. Also, from Euler's rotation theorem \cite{enwiki:1147410841}, we know that for two sets of arbitrary orthogonal vectors, the rotation matrix is unique. Since the eigenvectors of a two-by-two symmetric matrix are orthogonal, the RANS Reynolds stress eigenvectors can be mapped to the modified (true DNS) eigenvectors with a rotation around an axis and a certain angle in 3D. It should be noted that first the eigenvalues are arranged in order of magnitude and the eigenvectors such as each of these values are compared with the eigenvectors from the DNS simulation.
In general, for the rotation of a vector in three dimensions around a certain axis
$\mathbf{u} = (u_x,u_y,u_z)$
and the rotation angle $\theta$, the rotation matrix is considered as follows.
\begin{equation}
\resizebox{.9\hsize}{!}{$R=\left[\begin{array}{ccc}
\cos \theta+u_x^2(1-\cos \theta) & u_x u_y(1-\cos \theta)-u_z \sin \theta & u_x u_z(1-\cos \theta)+u_y \sin \theta \\
u_y u_x(1-\cos \theta)+u_z \sin \theta & \cos \theta+u_y^2(1-\cos \theta) & u_y u_z(1-\cos \theta)-u_x \sin \theta \\
u_z u_x(1-\cos \theta)-u_y \sin \theta & u_z u_y(1-\cos \theta)+u_x \sin \theta & \cos \theta+u_z^2(1-\cos \theta)
\end{array}\right]$}
\end{equation}
To maintain the rotational invariancy, the rotation matrix can be replaced with the rotation quadrature $\mathbf{q}$.
A rotation with angle $\theta$ around the axis defined by the unit vector
$\mathbf{u} = (u_x,u_y,u_z)$ can be represented by the following quaternion.
\begin{equation}
\mathbf{q} = \cos\frac{\theta}{2} + (u_x\mathbf{i},u_y\mathbf{j},u_z\mathbf{k})\sin\frac{\theta}{2 }
\end{equation}
It can be shown that the result of this period for an arbitrary vector
$\mathbf{p} = (p_x,p_y,p_z)$
 is determined by the following relation:
\begin{equation}\label{quaternion_rotation}
(0,\mathbf{p}') = \mathbf{q}(0,\mathbf{p})\mathbf{q}^{-1}
\end{equation}
which in the above relationship, the inverse of the quaternion 
$\mathbf{q}^{-1}$
is calculated according to the following relationship:
\begin{equation}
\mathbf{q}^{-1} = \cos\frac{\theta}{2} - (u_x\mathbf{i},u_y\mathbf{j},u_z\mathbf{k})\sin\frac{\theta}{2 }
\end{equation}
Now, by considering an eigenvector obtained from RANS simulation and assuming that the axis and rotation angle are known, the modified eigenvector (DNS) can be obtained.
According to Euler's rotation theorem, we know that the period matrix is unique for two sets of arbitrary orthogonal vectors. On the other hand, the eigenvectors of a two-by-two symmetric matrix are mutually orthogonal, and therefore only one axis and one rotation angle will determine the proper rotation of all three eigenvectors, which, taking into account the unity of the rotation axis, the variables $u_x$,
$u_y$
  and $\theta$ are added to the outputs of the transformation function.
\begin{itemize}
\item \textbf{Pre-processing on eigenvectors}\\
Since the negative of an eigenvector is also the eigenvector of an eigenvalue, and in the process of eigenvector calculation for RANS and DNS simulations, there is a possibility of encountering eigenvectors with an angle difference greater than $90^{\circ}$, it is necessary to present a process which, if necessary, replaced a eigenvector with its negative. After sorting the eigenvectors based on the magnitude of their corresponding eigenvalues, the inner product of the first eigenvectors obtained from RANS and DNS and the negative DNS is calculated and the product of the inner product specifies the correct direction of the desired eigenvector.
Also, to ensure that the group of eigenvectors is right-handed and considering the orthogonality of the eigenvectors in this problem, the third eigenvector is calculated from the outer product of the first two eigenvectors.
\end{itemize}
\subsection{Optimum value of turbulence viscosity}
In this section, a method for calculating the optimal value of the only remaining variable from the variables required to close the Eq. \eqref{RST_modeled}, i.e. $\nu_{t}$ is introduced. In fact, the goal is to obtain the turbulence viscosity to minimize the difference between the correct Reynolds stress and the Reynolds stress resulting from Boussinesq's hypothesis $-2\nu_{t}\mathbf{S}$.

Therefore, the desired optimization problem is according to the following relationship:
\begin{equation}
\nu_{t}^{L} = \arg \min_{\nu_t} ||R_{dev} + 2\nu_t S||
\end{equation}
The algebraic details of solving this optimization problem are given below.
$$
\frac{d}{d\nu_t} {||R_{dev} + 2\nu_t S||}^2 = 0 \xrightarrow{} \frac{d}{d\nu_t} {tr(R_{dev} R_{dev}^{T} + R_{dev}2\nu_t S^{T} + 2\nu_t S R_{dev}^{T} + 4\nu_{t}^2S S^T)} =0
$$

\begin{equation}\label{optimal_viscosity}
\nu_t = -1/2 \frac{R_{dev}:S}{||S||^ 2}
\end{equation}
in which
$R_{dev}:S$ indicates double dot production.

\subsection{ML networks used as RST mapping function}
Here is the improved version of your paragraphs, with grammar and flow enhancements while preserving the meaning and all LaTeX syntax:

In this research, multiple neural networks were employed to predict output variables corresponding to tensors $\Delta \tau$, $\Delta \tau^L$, and optimal turbulence viscosity $\nu_t$. Various machine learning architectures, including Multi-Layer Perceptrons (MLP), Convolutional Neural Networks (CNNs) utilizing DeepInsight to transform vector feature sets into 2D feature images, and Random Forests, were implemented. Mutual validation and grid search were conducted to fine-tune the hyperparameters effectively.

To ensure that the third component of the rotation axis vector is real, the following relationship must hold for the first two components:
$$
u_{x}^{2} + u_{y}^{2} \leq 1
$$
A correction term is added to the loss function to enforce this constraint. Typically, the correction term is designed as the sum of a least squares function for regression applications. The final form of the proposed cost function is given as follows:

\begin{equation}
Loss = MSE + \alpha(ReLU(u_{x}^{2}+u_{y}^{2}-1))
\end{equation}

The DeepInsight \cite{10.1038/s41598-019-47765-6} method is proposed to transform non-image samples into an organized image form. This transformation enables the use of Convolutional Neural Networks (CNNs), including GPU acceleration, for non-image datasets. While the order of features has no direct effect on methods such as random forests, decision trees, or MLPs, the reliability of these methods often depends on the feature extraction techniques employed.

In contrast, a CNN architecture accepts input as an image (i.e., a matrix of size $m \times n$) and performs feature extraction through hidden layers, such as convolutional layers, ReLU layers, and max-pooling layers. One significant advantage of this approach is its ability to uncover higher-order statistical features and nonlinear correlations within the data.

By considering the relationships among neighboring points, CNNs achieve a richer representation compared to traditional machine learning models that process points independently. The DeepInsight method leverages this capability by integrating three steps: element arrangement, feature extraction, and classification. It transforms data into images by grouping similar features close together and placing dissimilar features farther apart. This spatial arrangement enables the model to more effectively utilize contextual information, uncovering underlying patterns such as pathways or feature relationships that might otherwise remain hidden.

For instance, a feature vector $x$ is converted into a feature matrix $M$ through a transformation $T$. Each feature’s position in the image is determined by its similarity to other features in the dataset. This arrangement in Cartesian coordinates provides a visual representation that highlights the relationships among features.

\begin{figure}[h!]
\centering
\includegraphics[width=\textwidth]{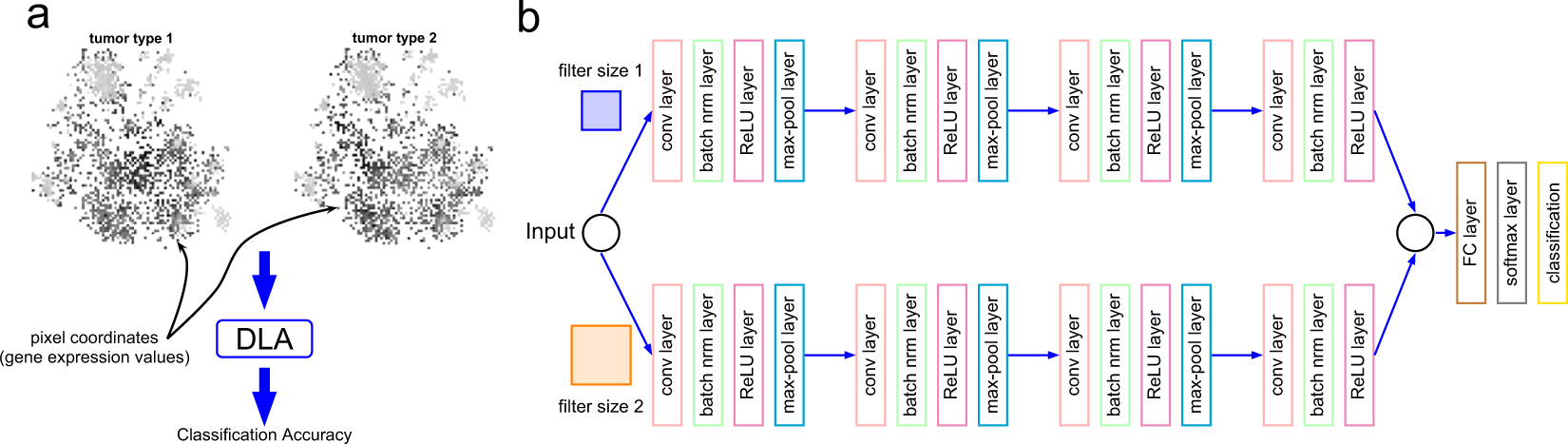}
\caption{Parallel architecture of convolutional neural network used for applying regression on output images from DeepInsight algorithm \cite{10.1038/s41598-019-47765-6}}
\label{fig:CNN}
\end{figure}

\section{Results and Discussion}
\subsection{Square Channel Flow}
First, both the training and the testing stages are applied to the fully developed, incompressible, and isothermal flow inside a channel with a square cross-section. The objective is to improve RANS simulations and achieve more accurate secondary flows predictions. An MLP is trained using DNS data and RANS simulations at a Reynolds number of $Re=2900$ to provide implicit and explicit terms in the modified equation \eqref{RST_eigen_decomposition}. This trained model is then applied to predict flow behavior at a higher Reynolds number of $Re=3500$.

\begin{figure}[ht]
\centering
\subfloat[secondary flow obtained from DNS simulation]{ \label{fig:Channel_sec_flow:DNS}
\includegraphics[width=0.33\textwidth]{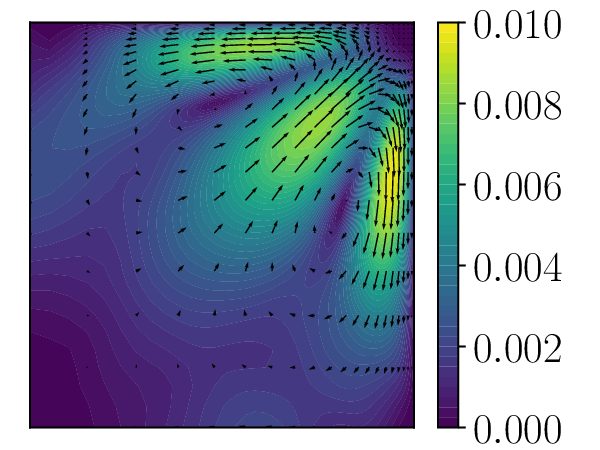}}
\subfloat[secondary flow obtained from RANS simulation with LRR model]{ \label{fig:Channel_sec_flow:LRR}
\includegraphics[width=0.33\textwidth]{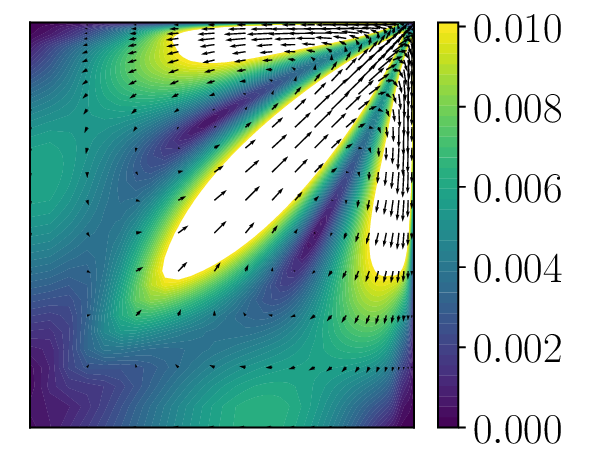}}%
\subfloat[Secondary flow obtained from RANS simulation with $k-\epsilon$ model]{ \label{fig:Channel_sec_flow:kep}
\includegraphics[width=0.33\textwidth]{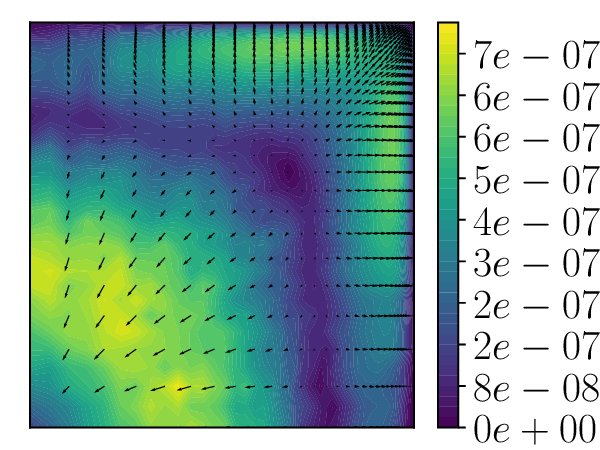}}%

\caption {Secondary flow in a channel with a square section (numbers in colored columns indicate the size of the secondary velocity vector) }
\label{fig:Channel_sec_flow} 
\end{figure}

The results of the RANS simulation for the secondary flow in the channel with a square cross section, obtained using two models $k-\epsilon$ and LRR are shown alongside the DNS reference data in Fig. \ref{fig:Channel_sec_flow:DNS}. It is evident that the $k-\epsilon$ model fails to accurately predict the secondary currents, and the results from the LRR model also deviate significantly from the DNS reference data.

\begin{figure}[h!]
  \centering
  \subfloat[]{ \label{fig:RANS_inputs_sample:1}
  \includegraphics[width=0.33\textwidth]{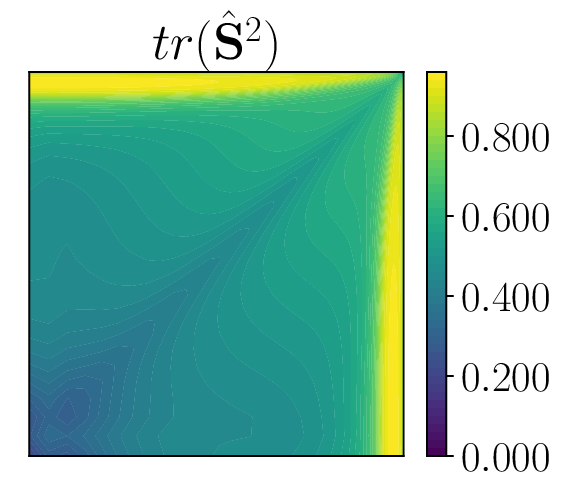}}
  \subfloat[]{ \label{fig:RANS_inputs_sample:2}
  \includegraphics[width=0.33\textwidth]{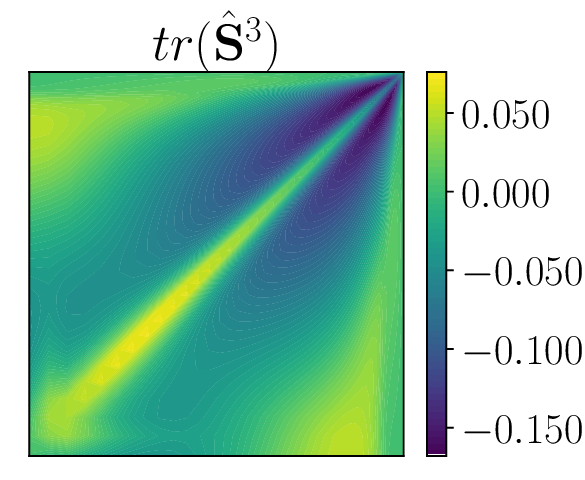}}%
  \subfloat[]{ \label{fig:RANS_inputs_sample:3}
  \includegraphics[width=0.33\textwidth]{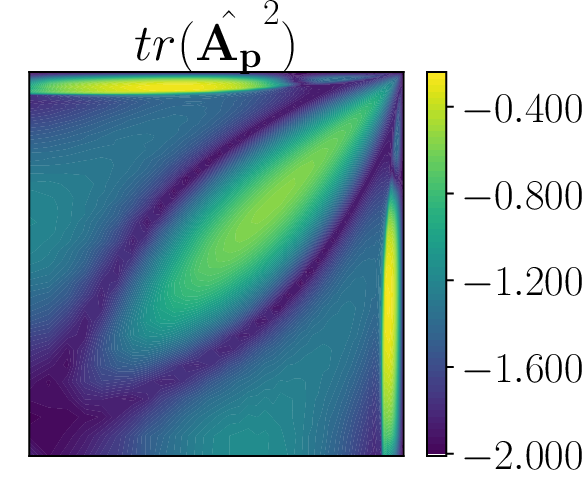}}%
 
  \hfill
  \subfloat[]{ \label{fig:RANS_inputs_sample:4}
\includegraphics[width=0.33\textwidth]{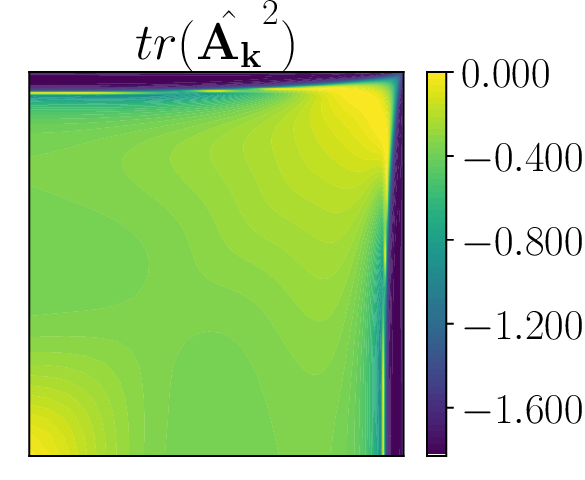}}
\subfloat[]{ \label{fig:RANS_inputs_sample:5}
\includegraphics[width=0.33\textwidth]{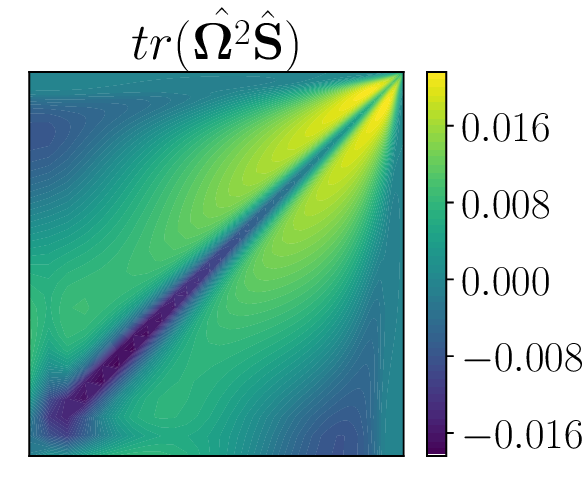}}%
\subfloat[]{ \label{fig:RANS_inputs_sample:6}
\includegraphics[width=0.33\textwidth]{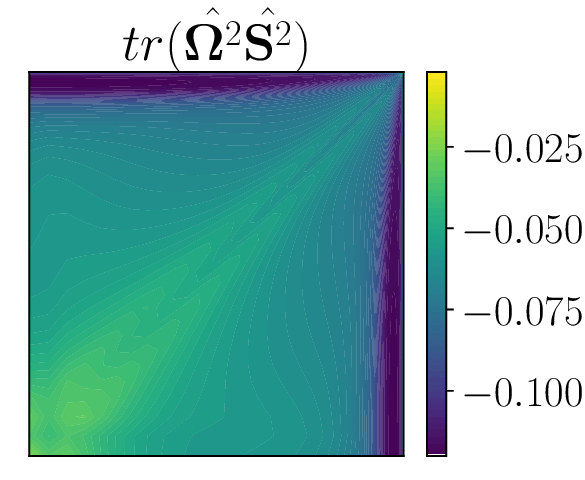}}%

  \caption{The first 6 inputs of the neural network obtained from the converged RANS solution}
  \label{fig:RANS_inputs_sample} 
\end{figure}

According to the previous section, from the 4 basic tensors i.e.
$\mathbf{Q} = \{\mathbf{S},\mathbf{\Omega}, A_p, A_k\}$
and 3 scalars which all are obtained from simulation RANS, 47 secondary invariants are obtained. The trace of the first six matrices from the set of 47 matrices, which are 
$\hat{\mathbf{S}}^2$
,
$\hat{\mathbf{S}}^3$
,
$\hat{\mathbf{A}_p}^2$
,
$\hat{\mathbf{A}_k}^2$
,
$\hat{\mathbf{\Omega}}^2\hat{S}$
and
$\hat{\mathbf{\Omega}}^2\hat{S}^2$
are shown in Figs \ref{fig:RANS_inputs_sample:1} -\ref{fig:RANS_inputs_sample:6}.

\begin{figure}[h!]
\centering
\subfloat[]{ \label{fig:DeltaEigenValue:xi}
\includegraphics[width=0.33\textwidth]{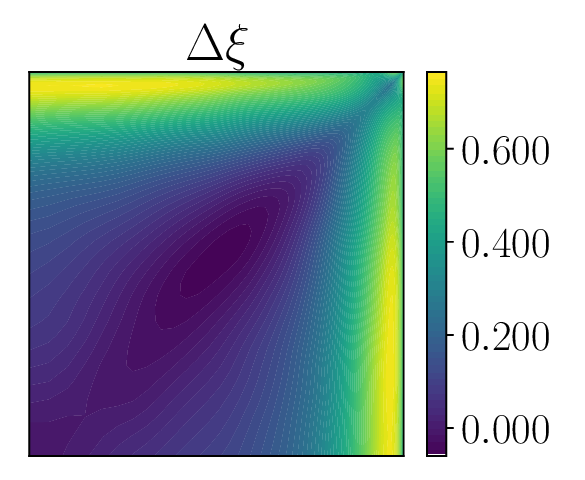}}
\subfloat[]{ \label{fig:DeltaEigenValue:eta}
\includegraphics[width=0.33\textwidth]{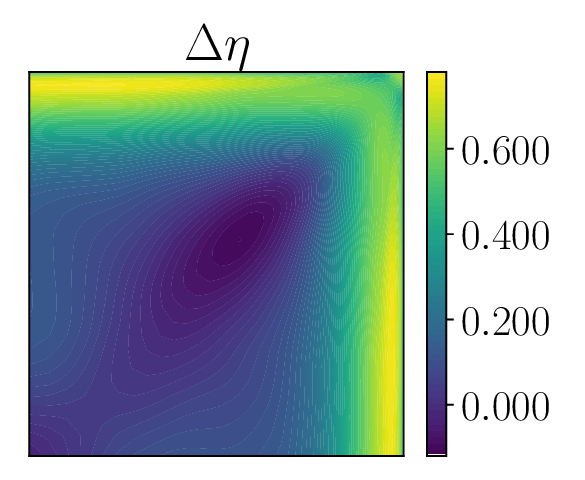}}%
\caption {Difference of eigenvalues obtained from the deviatoric part of Reynolds stress of DNS and RANS data}
\label{fig:DeltaEigenValue} 
\end{figure}

The difference of the eigenvalues mapped on the barycentric coordinate system can be calculated first to construct a set of output variables. Figs
\ref{fig:DeltaEigenValue:xi}
  and
\ref{fig:DeltaEigenValue:eta}
   are indicating this difference.

\begin{figure}[h!]
\centering
\subfloat[]{ \label{fig:theta:x}
\includegraphics[width=0.33\textwidth]{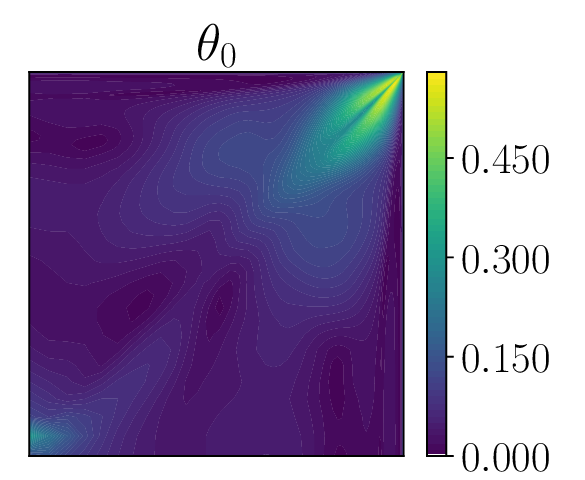}}
\subfloat[]{ \label{fig:theta:y}
\includegraphics[width=0.33\textwidth]{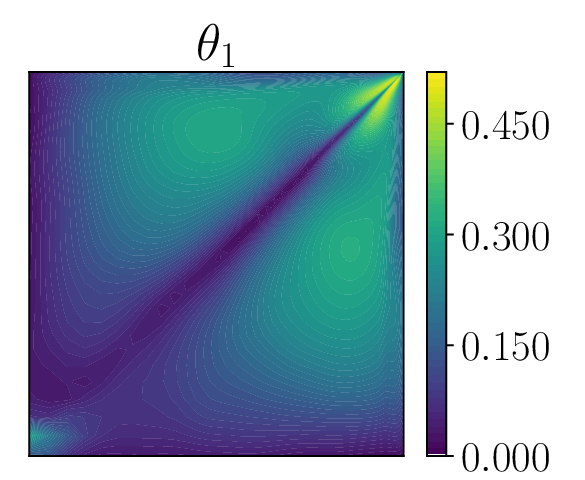}}%
\subfloat[]{ \label{fig:theta:z}
\includegraphics[width=0.33\textwidth]{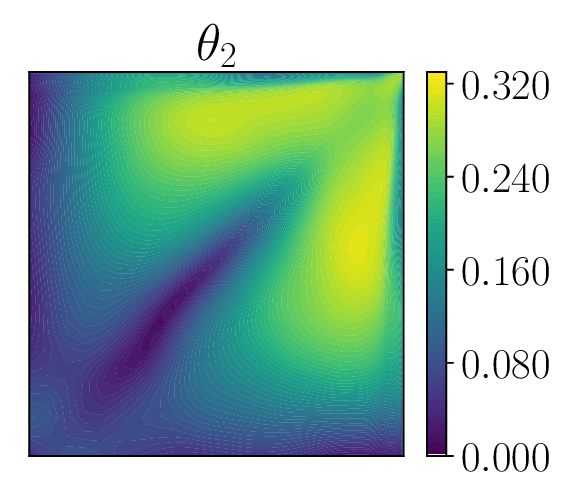}}%
\caption {Suitable rotation angles for converting eigenvectors obtained from RANS to DNS}
\label{fig:theta} 
\end{figure}

\begin{figure}[h!]
\centering
\subfloat[]{ \label{fig:axis:x}
\includegraphics[width=0.33\textwidth]{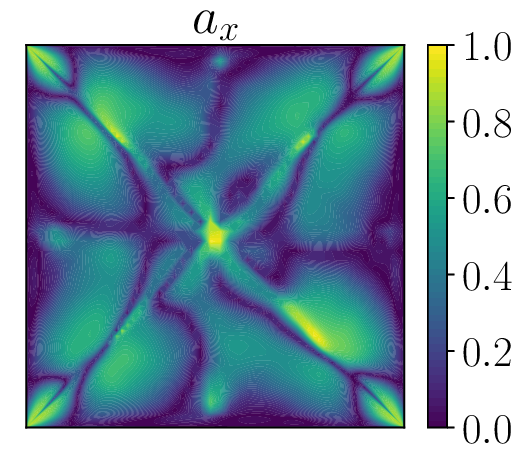}}
\subfloat[]{ \label{fig:axis:y}
\includegraphics[width=0.33\textwidth]{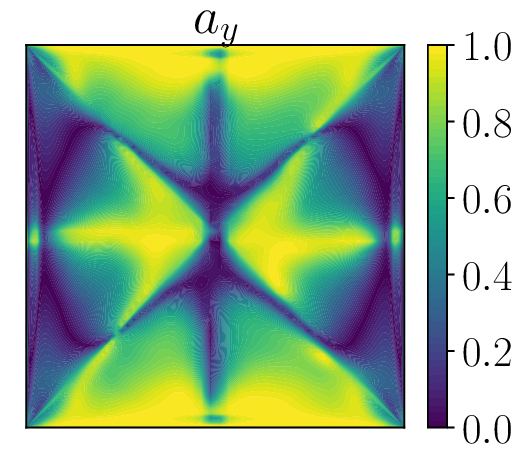}}%
\subfloat[]{ \label{fig:axis:z}
\includegraphics[width=0.33\textwidth]{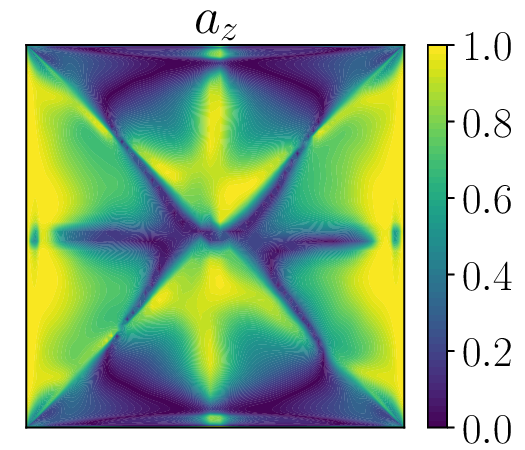}}%

\caption {The appropriate rotation axis components for converting eigenvectors obtained from RANS to DNS}
\label{fig:axis} 
\end{figure}

Also, for better intuition and understanding, the eigenvectors of deviatoric Reynolds stress have been compared for the point
$y,z=0$. It is known that for each simulation mode, due to the symmetry of the deviatoric Reynolds stress matrix, the eigenvectors are orthogonal, which can be seen in the figure below.
According to the relation \ref{quaternion_rotation}, the angle and axis of rotation necessary to convert the feature vectors obtained from RANS simulation to DNS can be calculated, which are shown in Figs.
\ref{fig:theta}
  and
\ref{fig:axis}
   are shown.

\begin{figure}[h!]
\centering
\subfloat[Cost function diagram for network with non-axis outputs]{ \label{fig:MLP_Loss:non_axis}
\includegraphics[width=0.4\textwidth]{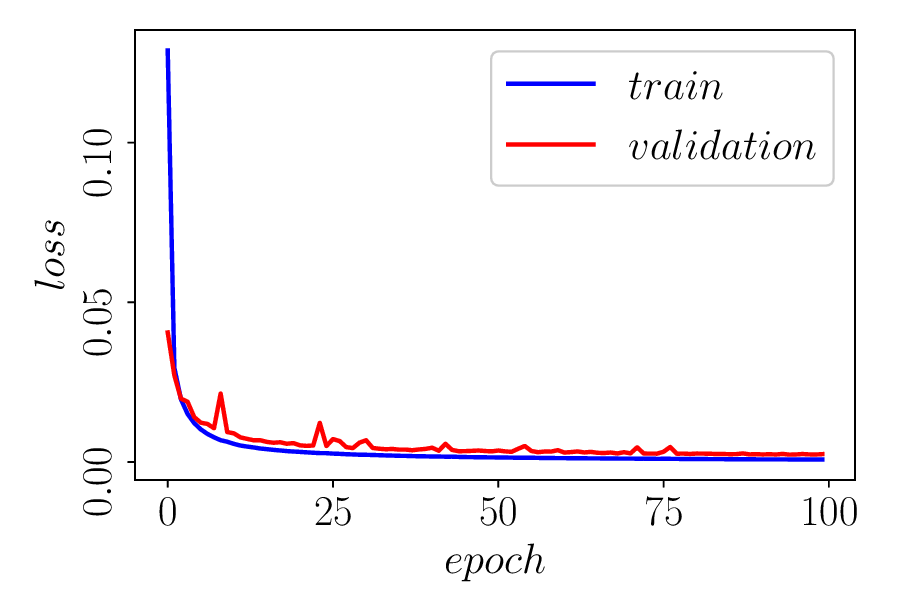}}
\subfloat[Cost function diagram for the network with the outputs of the rotation axis elements]{ \label{fig:MLP_Loss:axis}
\includegraphics[width=0.4\textwidth]{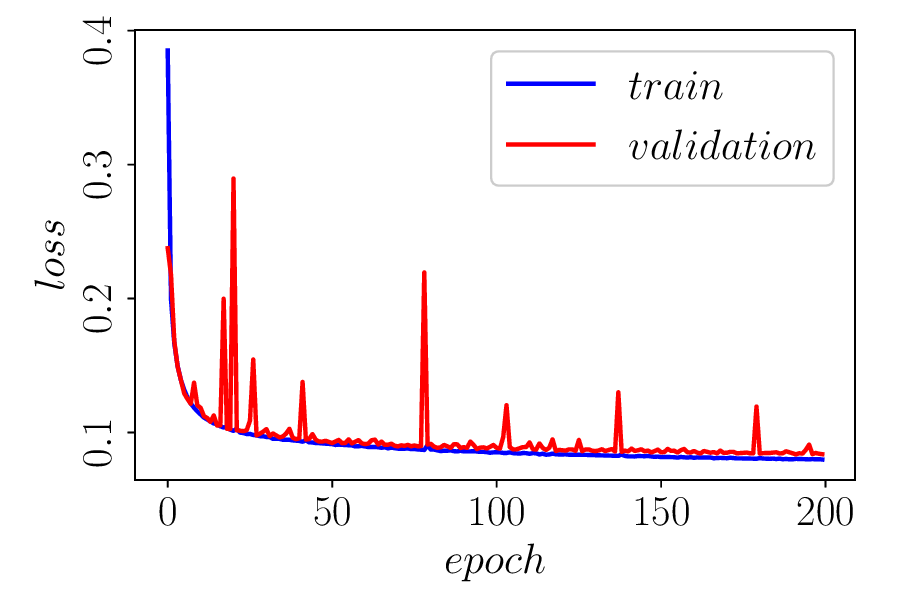}}%
\caption {Cost functions for learning data and validation outputs for comparing RANS with DNS}
\label{fig:MLP_Loss} 
\end{figure}

A multi-layer perceptron neural network was chosen for this problem in which the number of neurons of different layers and the activation functions are given in tables
\ref{tab:MLP_specs}
and
\ref{tab:MLP_axis_specs}.
In Fig. \ref{fig:MLP_Loss} graphs of the cost function are drawn.
\begin{table}[ht]
\caption {Perceptron neural network specifications for all outputs except rotation axes}
\label{tab:MLP_specs}
\centering
\onehalfspacing
\begin{tabular}{|c|c|c|}
\hline layer & number of neurons & activation function \\
\hline 1 & 50 & $ReLU$ \\
\hline 2 & 50 & $ReLU$ \\
\hline 3 & 50 & $ReLU$ \\
\hline 4 & 11 & $Linear$ \\
\hline optimizer & \multicolumn{2}{c|}{SGD} \\
\hline cost function & \multicolumn{2}{c|}{MSE} \\
\hline
\end{tabular}
\end{table}

\begin{table}[ht]
\caption {Perceptron neural network specifications for elements of rotation axes}
\label{tab:MLP_axis_specs}
\centering
\onehalfspacing
\begin{tabular}{|c|c|c|}
\hline layer & number of neurons & activation function \\
\hline 1 & 50 & $ReLU$ \\
\hline 2 & 50 & $ReLU$ \\
\hline 3 & 50 & $ReLU$ \\
\hline 4 & 6 & $Linear$ \\
\hline optimizer & \multicolumn{2}{c|}{SGD} \\
\hline cost function & \multicolumn{2}{c|}{MSE+Regularization Term} \\
\hline
\end{tabular}
\end{table}

\begin{figure}[h!]
\centering
\subfloat[$\overline{u_y}$ at the section $y/h=0.25$]{ \label{fig:Comparison_Re2900_3500:sec25}
\includegraphics[width=0.33\textwidth]{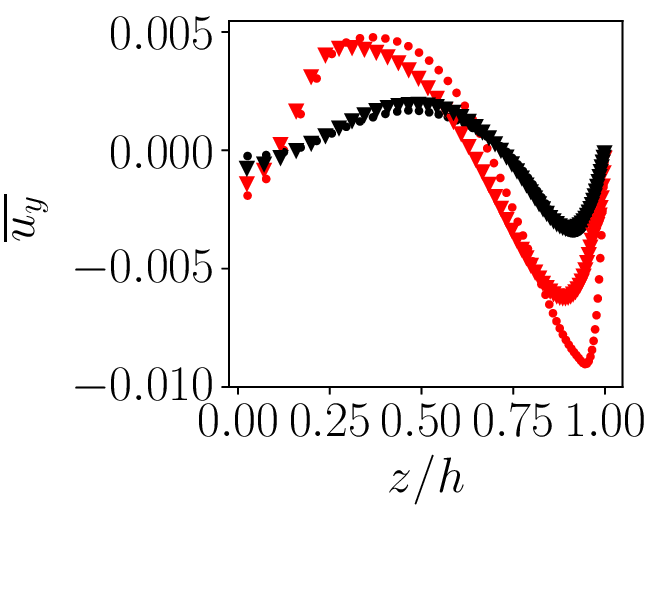}}
\subfloat[$\overline{u_y}$ at the section $y/h=0.5$]{ \label{fig:Comparison_Re2900_3500:sec50}
\includegraphics[width=0.33\textwidth]{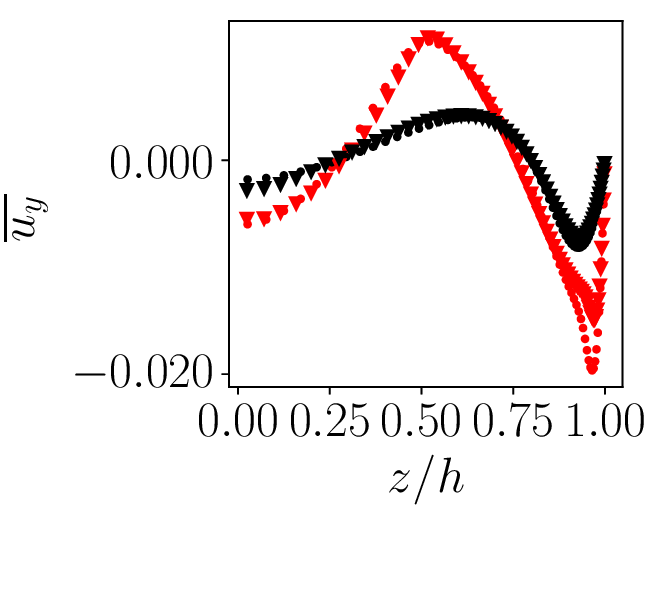}}%
\subfloat[$\overline{u_y}$ at cross section $y/h=0.75$]{ \label{fig:Comparison_Re2900_3500:sec75}
\includegraphics[width=0.33\textwidth]{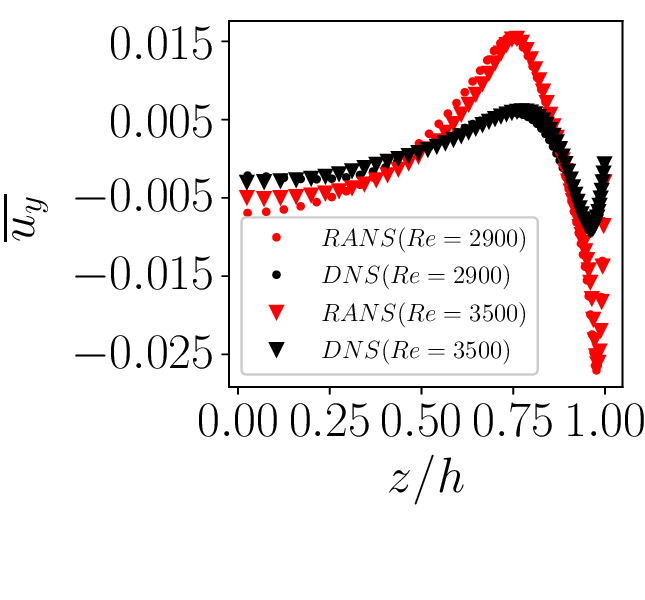}}%

\caption{Comparison of secondary flow
$\overline{u_y}$
obtained from RANS simulation and DNS data for two Reynolds 2900 and 3500}
\label{fig:Comparison_Re2900_3500} 
\end{figure}

\begin{figure}[h!]
\centering
\subfloat[$\overline{u_y}$ at cross section $y/h=0.25$]{ \label{fig:u_y_Re2900:sec25}
\includegraphics[width=0.33\textwidth]{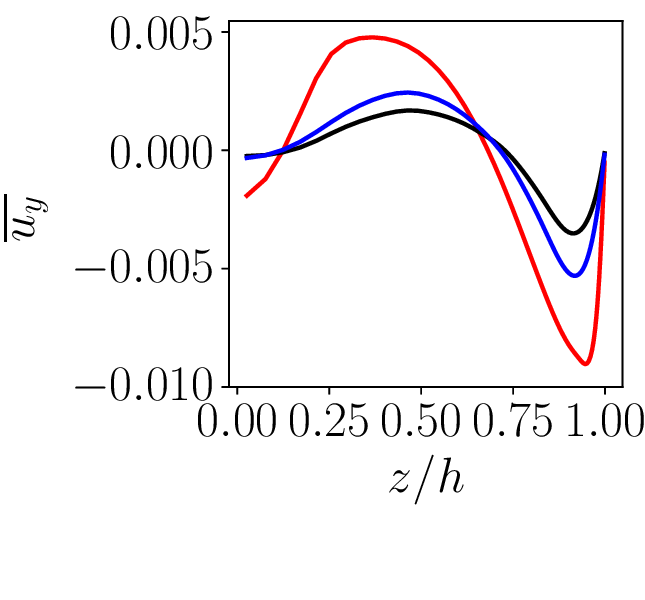}}
\subfloat[$\overline{u_y}$ at cross section $y/h=0.5$]{ \label{fig:u_y_Re2900:sec50}
\includegraphics[width=0.33\textwidth]{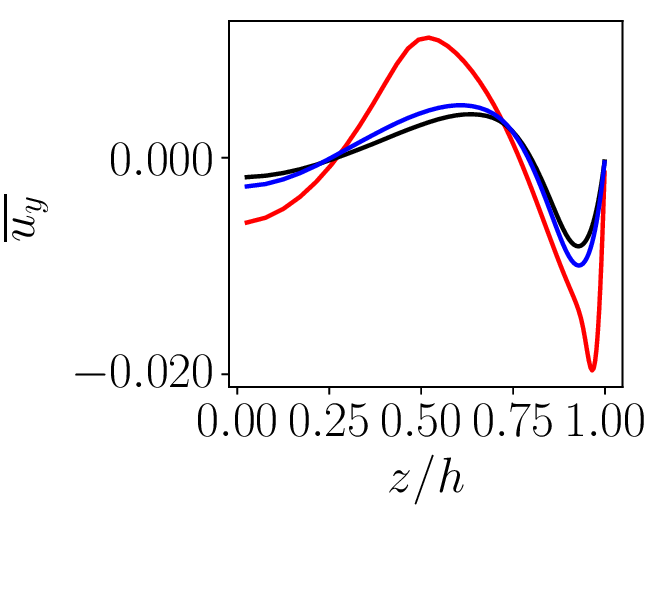}}%
\subfloat[$\overline{u_y}$ at cross section $y/h=0.75$]{ \label{fig:u_y_Re2900:sec75}
\includegraphics[width=0.33\textwidth]{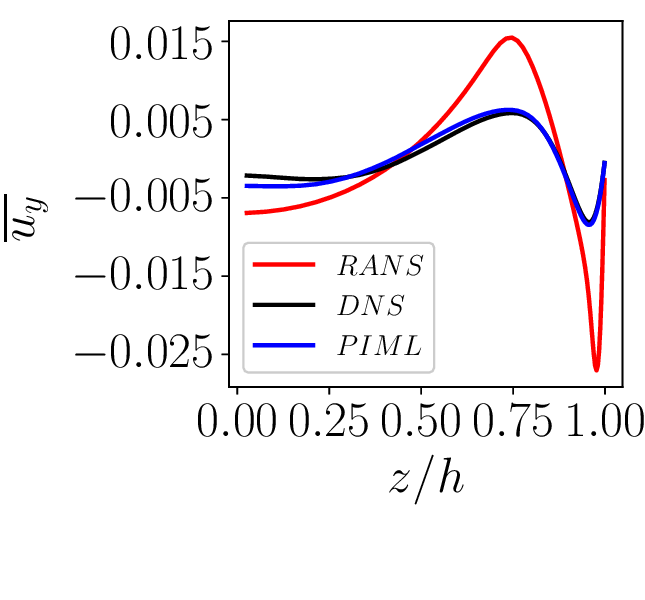}}%

\caption{Comparison of secondary flow
$\overline{u_y}$
obtained from RANS simulation, DNS data and PIML simulation for Re=2900}
\label{fig:u_y_Re2900} 
\end{figure}

\begin{figure}[h!]
\centering
\subfloat[$\overline{u_z}$ at cross section $y/h=0.25$]{ \label{fig:u_z_Re2900:sec25}
\includegraphics[width=0.33\textwidth]{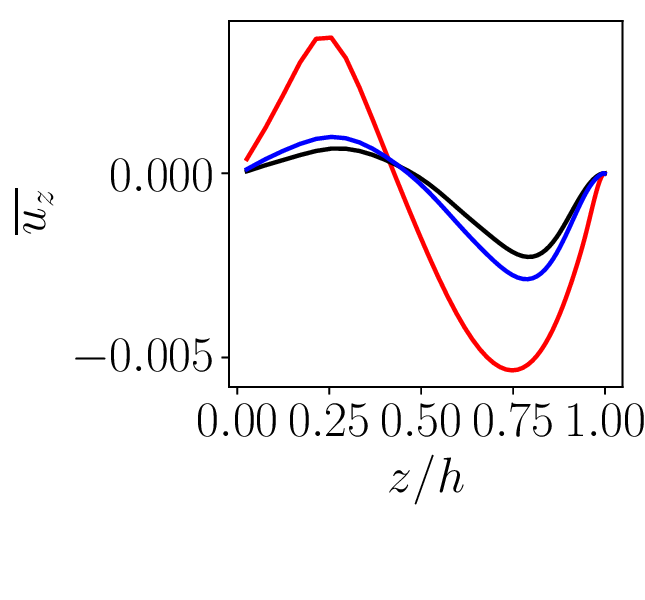}}
\subfloat[$\overline{u_z}$ at cross section $y/h=0.5$]{ \label{fig:u_z_Re2900:sec50}
\includegraphics[width=0.33\textwidth]{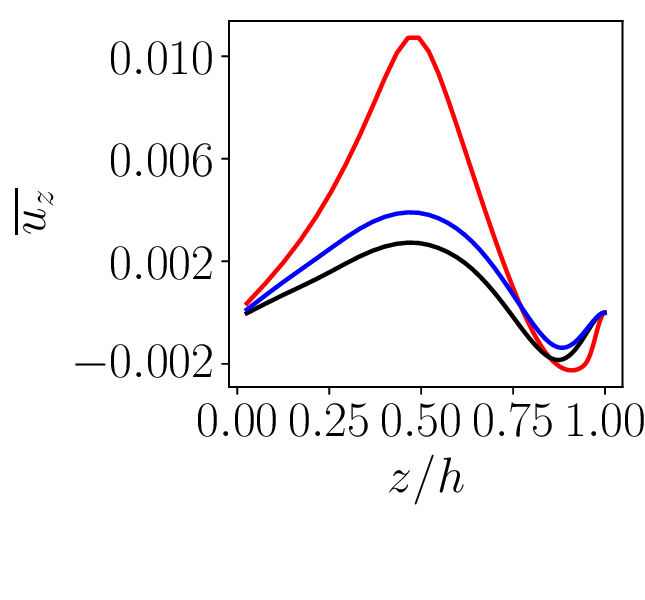}}%
\subfloat[$\overline{u_z}$ at cross section $y/h=0.75$]{ \label{fig:u_z_Re2900:sec75}
\includegraphics[width=0.33\textwidth]{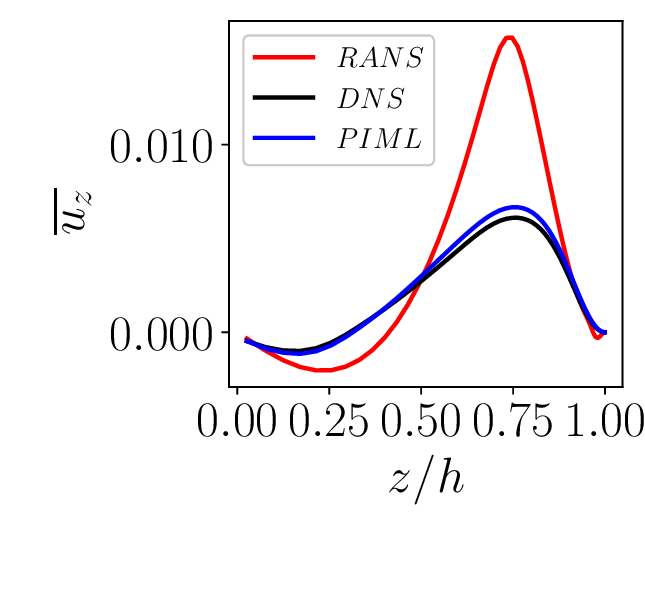}}%

\caption{Comparison of secondary flow
$\overline{u_z}$
obtained from RANS simulation, DNS data and PIML simulation for Re=2900}
\label{fig:u_z_Re2900} 
\end{figure}

The results of applying the algorithm with the trained MLP on $Re = 2900$ will be utilized for both $Re = 2900$, and $Re = 3500$. Velocity profiles obtained from RANS and DNS simulation are compared for both Reynolds numbers in the Figs. \ref{fig:Comparison_Re2900_3500}

. The effectiveness of the PIML method is assessed at the lower Reynolds number ($Re = 2900$) by evaluating its corrections to the secondary flow velocities $\overline{u_y}$ and $\overline{u_z}$, compared to results from RANS and DNS (Figures \ref{fig:u_y_Re2900} and \ref{fig:u_z_Re2900}).

\begin{figure}[h!]
\centering
\subfloat[$\overline{u_y}$ at cross section $y/h=0.25$]{ \label{fig:u_y_Re3500:sec25}
\includegraphics[width=0.33\textwidth]{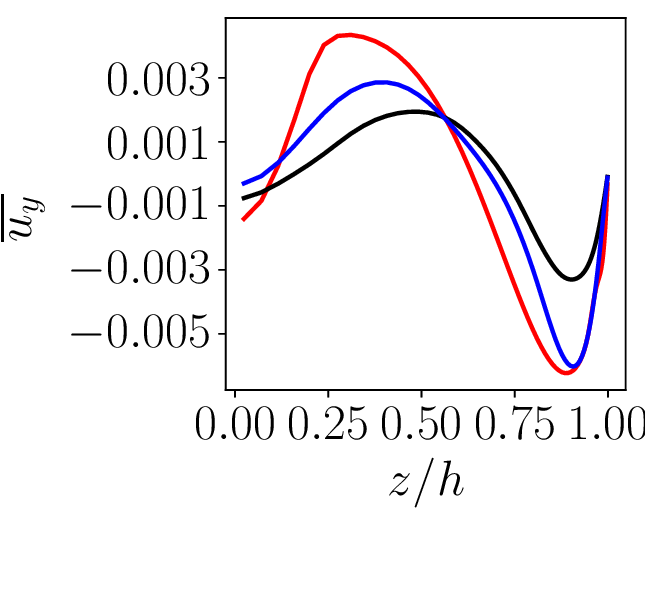}}
\subfloat[$\overline{u_y}$ at cross section $y/h=0.5$]{ \label{fig:u_y_Re3500:sec50}
\includegraphics[width=0.33\textwidth]{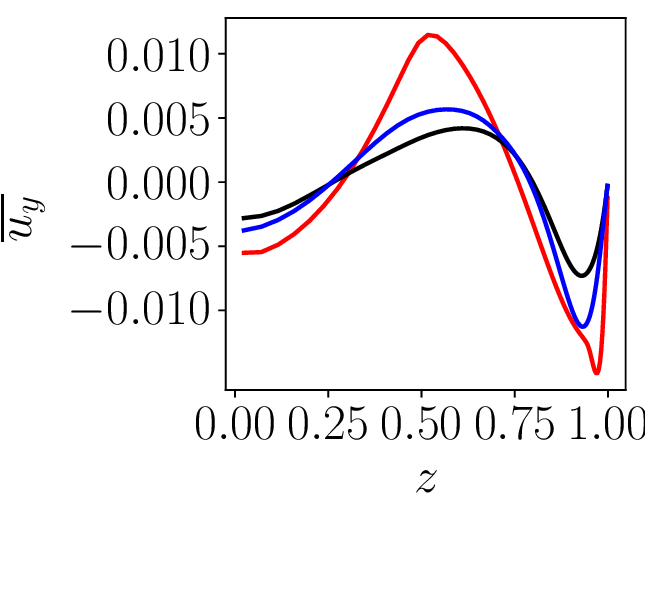}}%
\subfloat[$\overline{u_y}$ at cross section $y/h=0.75$]{ \label{fig:u_y_Re3500:sec75}
\includegraphics[width=0.33\textwidth]{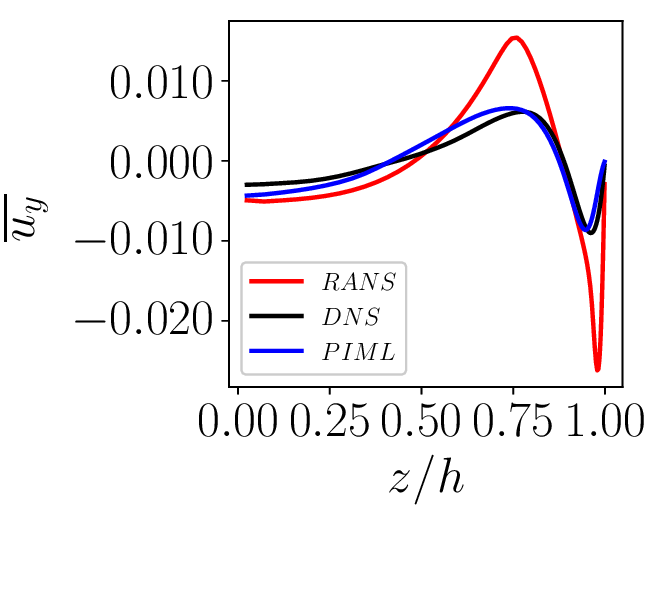}}%

\caption{Comparison of secondary flow
$\overline{u_y}$
Obtained from RANS simulation, DNS data and PIML simulation for Reynolds 3500}
\label{fig:u_y_Re3500} 
\end{figure}

\begin{figure}[h!]
\centering
\subfloat[$\overline{u_z}$ at cross section $y/h=0.25$]{ \label{fig:u_z_Re3500:sec25}
\includegraphics[width=0.33\textwidth]{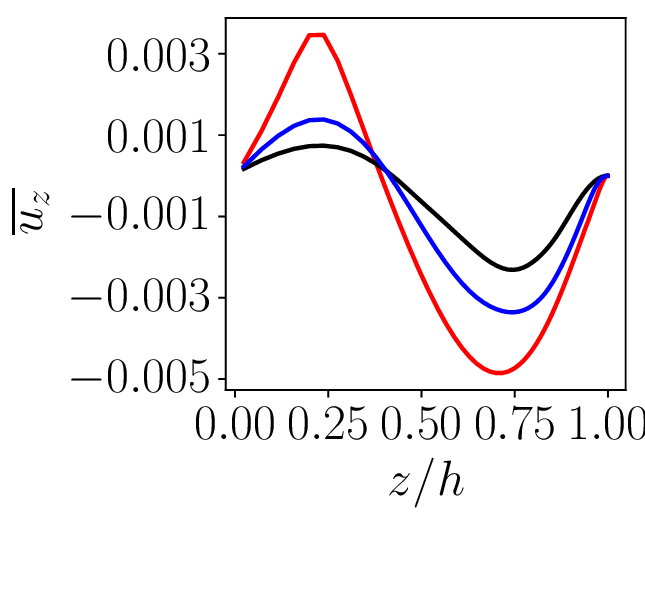}}
\subfloat[$\overline{u_z}$ at cross section $y/h=0.5$]{ \label{fig:u_z_Re3500:sec50}
\includegraphics[width=0.33\textwidth]{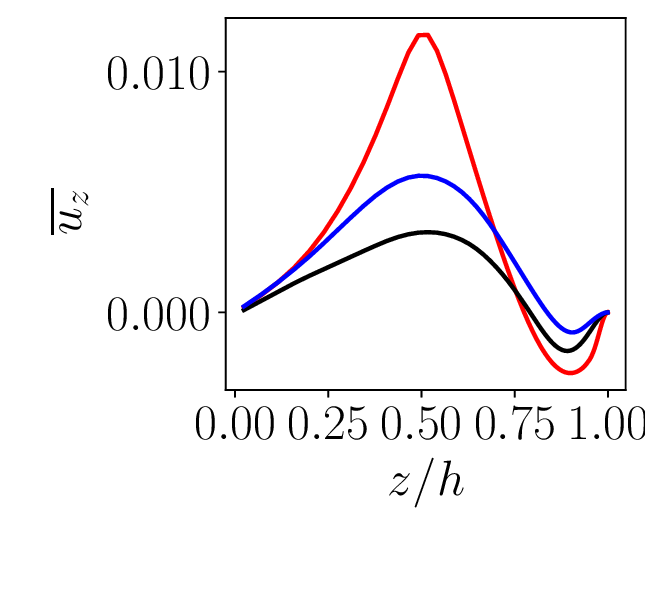}}%
\subfloat[$\overline{u_z}$ at cross section $y/h=0.75$]{ \label{fig:u_z_Re3500:sec75}
\includegraphics[width=0.33\textwidth]{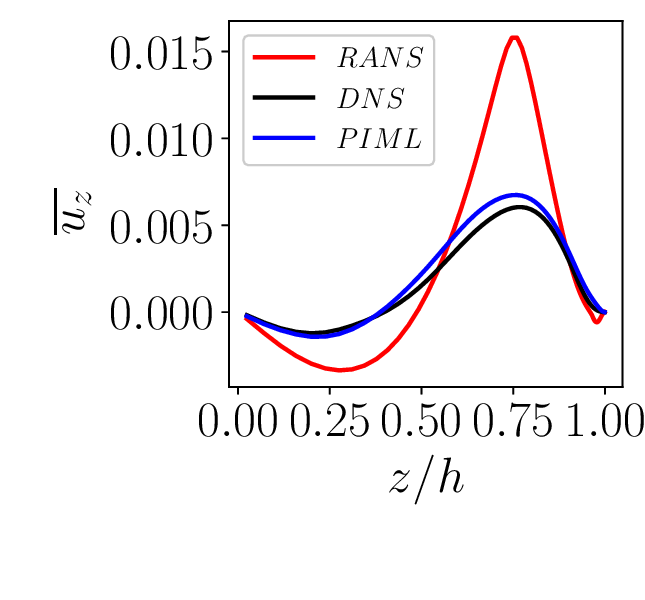}}%

\caption{Comparison of secondary flow
$\overline{u_z}$
obtained from RANS simulation, DNS data and PIML simulation for Re=3500}
\label{fig:u_z_Re3500} 
\end{figure}

\begin{figure}[h!]
\centering
\subfloat[$\overline{u_y}$ at cross section $y/h=0.25$]{ \label{fig:u_y_Re3500_convergence:sec25}
\includegraphics[width=0.33\textwidth]{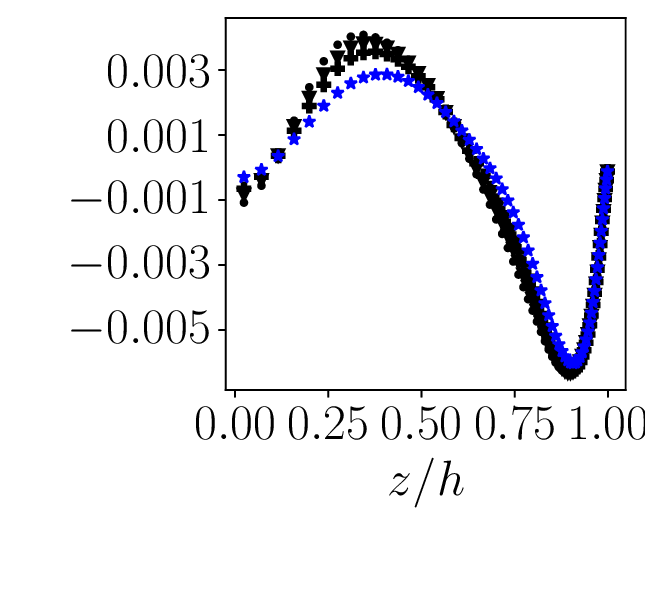}}
\subfloat[$\overline{u_y}$ at cross section $y/h=0.5$]{ \label{fig:u_y_Re3500_convergence:sec50}
\includegraphics[width=0.33\textwidth]{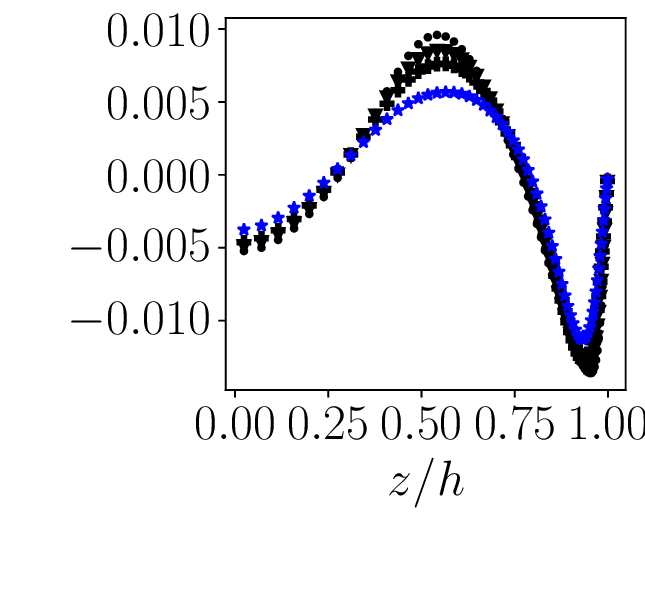}}%
\subfloat[$\overline{u_y}$ at cross section $y/h=0.75$]{ \label{fig:u_y_Re3500_convergence:sec75}
\includegraphics[width=0.33\textwidth]{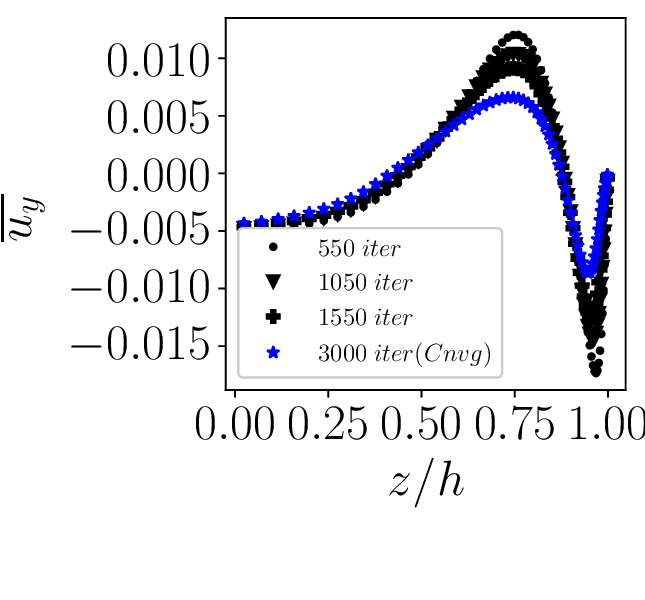}}%

\caption{ Secondary flow
$\overline{u_y}$
for Reynolds 3500 after successive iterations until convergence}
\label{fig:u_y_Re3500_convergence} 
\end{figure}

The velocity profiles as a function of the distance from the wall for $Re = 3500$ are shown in Figs. \ref{fig:u_y_Re3500} and \ref{fig:u_z_Re3500}. The modified PIML equation was solved using terms predicted by the MLP trained on $Re = 2900$. Additionally, the convergence and changes across successive iterations are illustrated in Fig. \ref{fig:u_y_Re3500_convergence} for $\overline{u_y}$. It is important to note that in each iteration, the simulation outputs are fed back as inputs to the MLP, updating the correction terms $\tau$, $\tau_{L}$, and $\nu_{t}^{L}$.

\begin{figure}[h!]
\centering
\subfloat[component $\tau_{yy}$
Reynolds stress in three sections
$y=0.25,0.50,0.75$
]{ \label{fig:tau_nu_comparison:yy}
\includegraphics[width=0.33\textwidth]{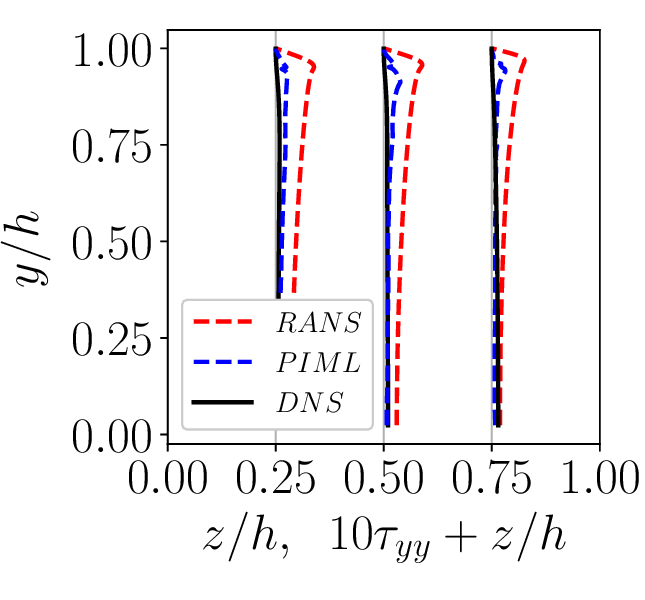}}
\subfloat[component $\tau_{zz}$
Reynolds stress in three sections
$y=0.25,0.50,0.75$
]{ \label{fig:tau_nu_comparison:zz}
\includegraphics[width=0.33\textwidth]{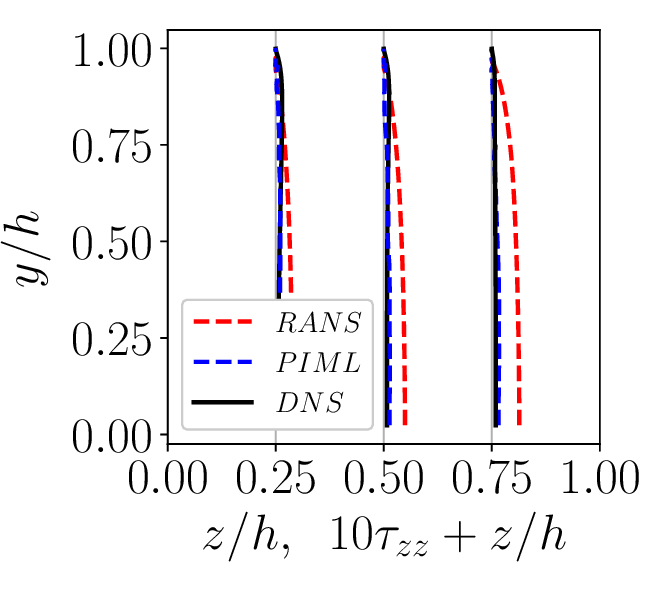}}%
\subfloat[optimal turbulence viscosity
$\nu_t$
in three stages
$y=0.25,0.50,0.75$
]{ \label{fig:tau_nu_comparison:nu}
\includegraphics[width=0.33\textwidth]{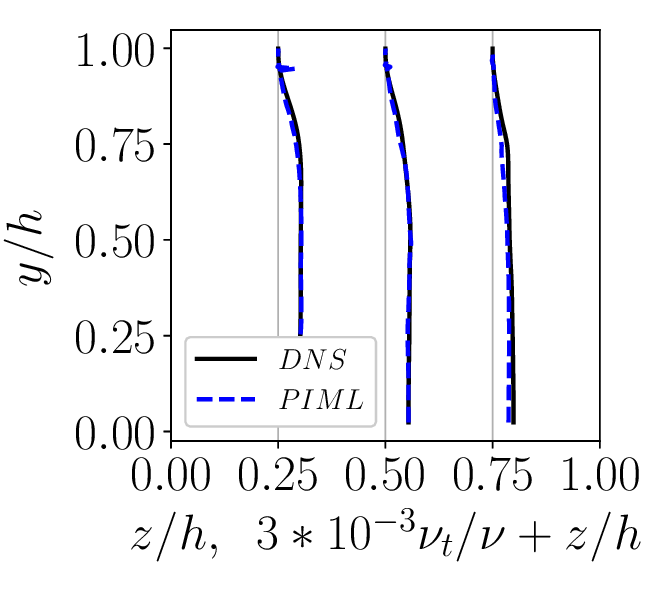}}%

\caption {Comparison of normal components in the secondary flow plane of Reynolds stress and optimal turbulence viscosity}
\label{fig:tau_nu_comparison} 
\end{figure}

Figures \ref{fig:tau_nu_comparison:yy} and \ref{fig:tau_nu_comparison:zz} compare the $\tau_{yy}$ and $\tau_{zz}$ components of the Reynolds stress tensor obtained from PIML, RANS, and DNS across three stages. The graphs indicate that the Reynolds stress predicted by the machine learning-modified equation for both components moves closer to the DNS data.


\begin{figure}[h]
\centerline{\includegraphics[width=0.35\textwidth]{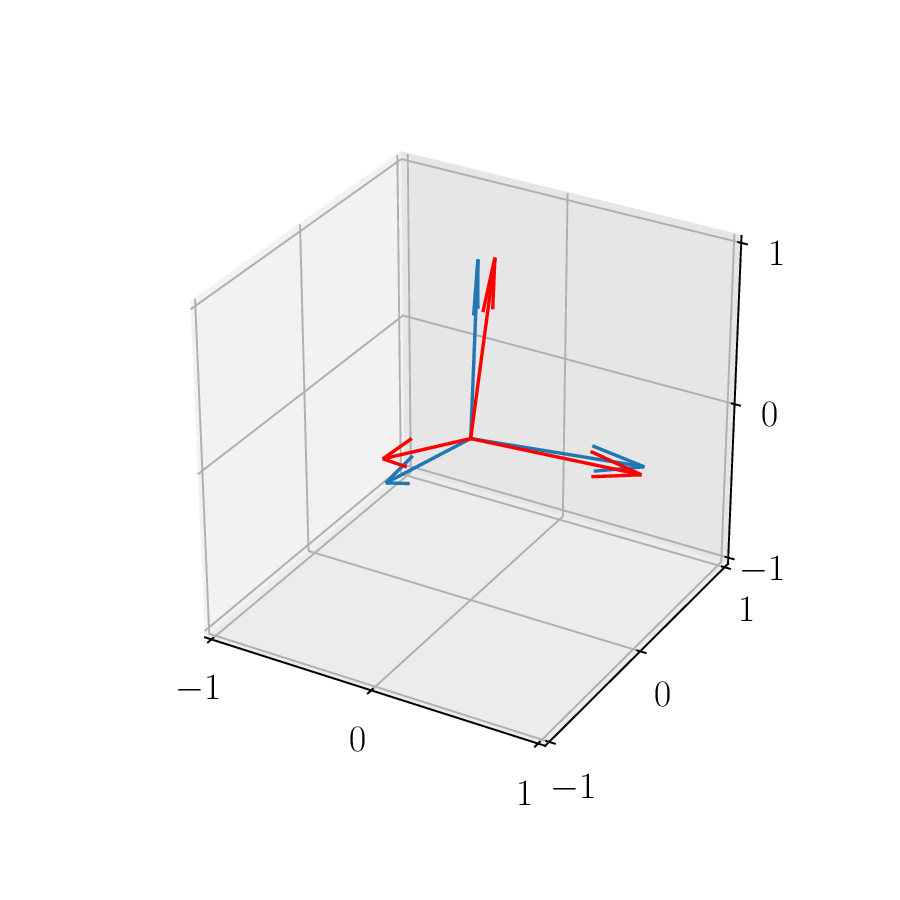}}
\caption {Unit and orthogonal eigenvectors of deviatoric Reynolds stress tensor}
\label{fig:EigenVectors}
\end{figure}
\begin{figure}[h]
\centerline{\includegraphics[width=0.35\textwidth]{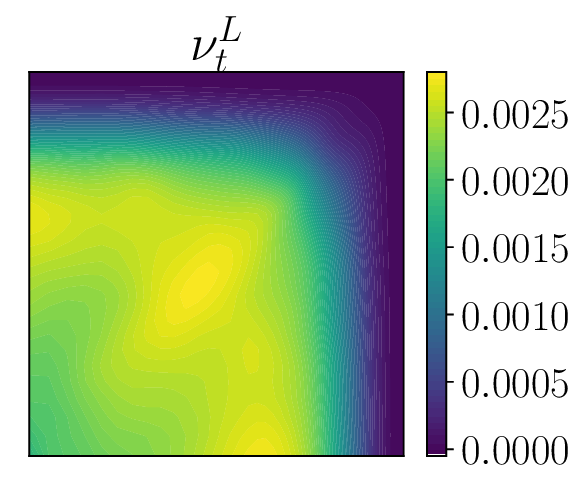}}
\caption {Optimum turbulence viscosity obtained from DNS data}
\label{fig:nu_t_L_DNS}
\end{figure}

\begin{figure}[h!]
\centering
\subfloat[Cost function diagram for network with non-axis outputs]{ \label{fig:MLP_Loss_L:non_axis}
\includegraphics[width=0.4\textwidth]{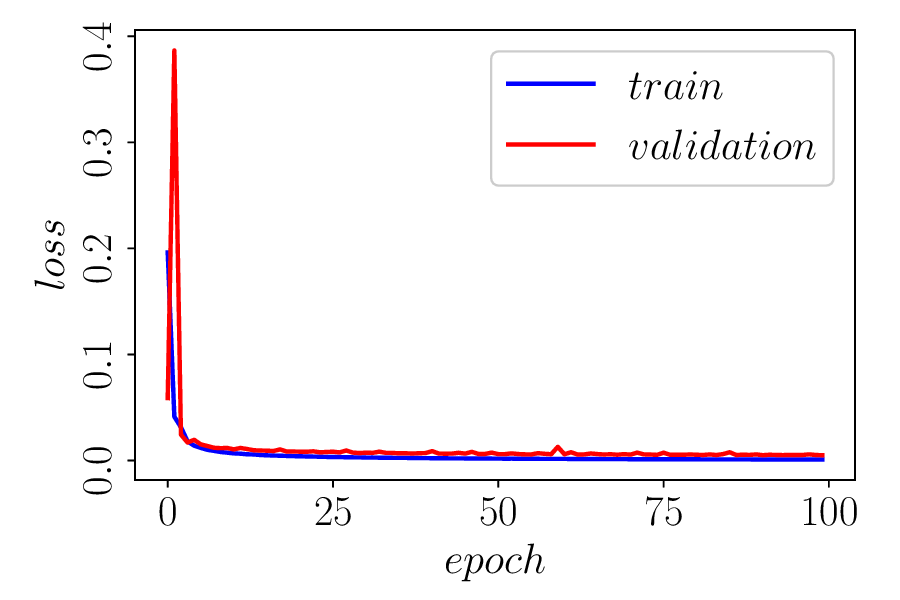}}
\subfloat[Cost function diagram for the network with the outputs of the rotation axis elements]{ \label{fig:MLP_Loss_L:axis}
\includegraphics[width=0.4\textwidth]{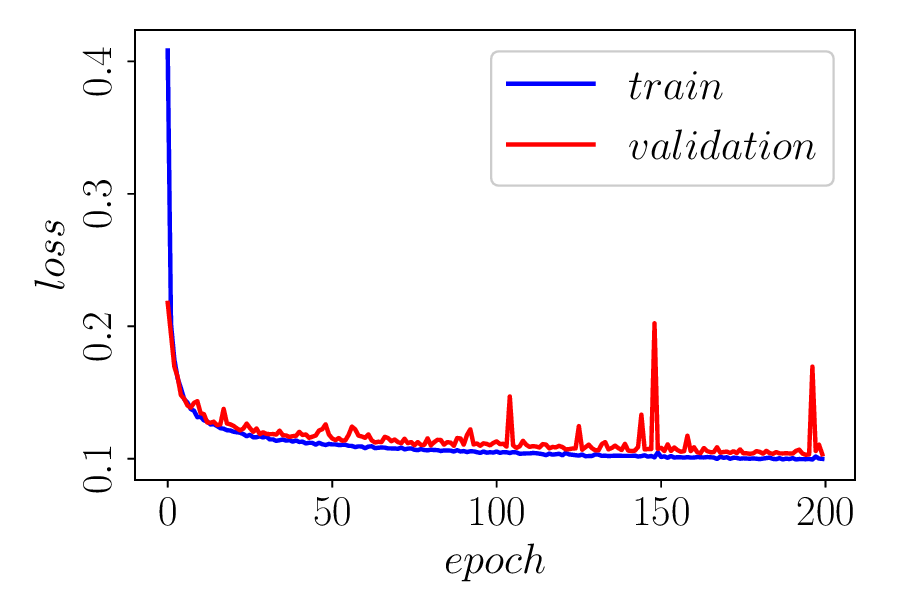}}%
\caption{Cost functions for learning data and validation outputs for comparing RANS with the linear part of DNS}
\label{fig:MLP_Loss_L} 
\end{figure}


\begin{figure}[h!]
\centering
\subfloat[$\overline{u_y}$ at cross section $y/h=0.25$]{ \label{fig:nu_t_Re3500:sec25}
\includegraphics[width=0.33\textwidth]{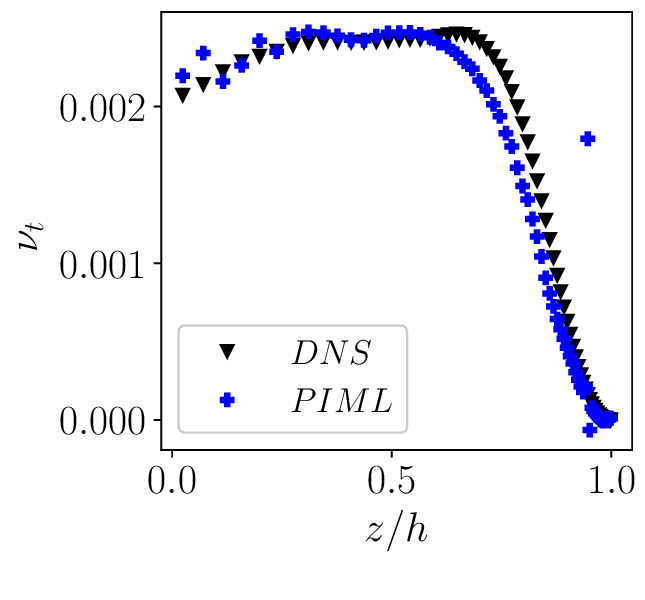}}
\subfloat[$\overline{u_y}$ at cross section $y/h=0.5$]{ \label{fig:nu_t_Re3500:sec50}
\includegraphics[width=0.33\textwidth]{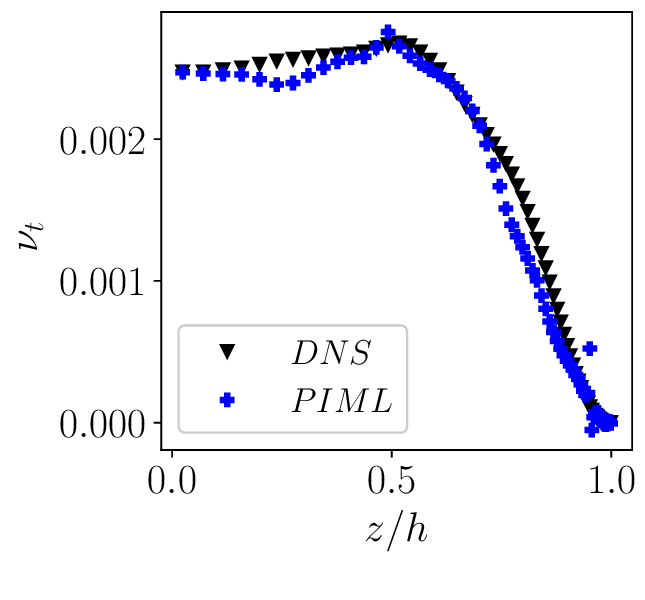}}%
\subfloat[$\overline{u_y}$ at cross section $y/h=0.75$]{ \label{fig:nu_t_Re3500:sec75}
\includegraphics[width=0.33\textwidth]{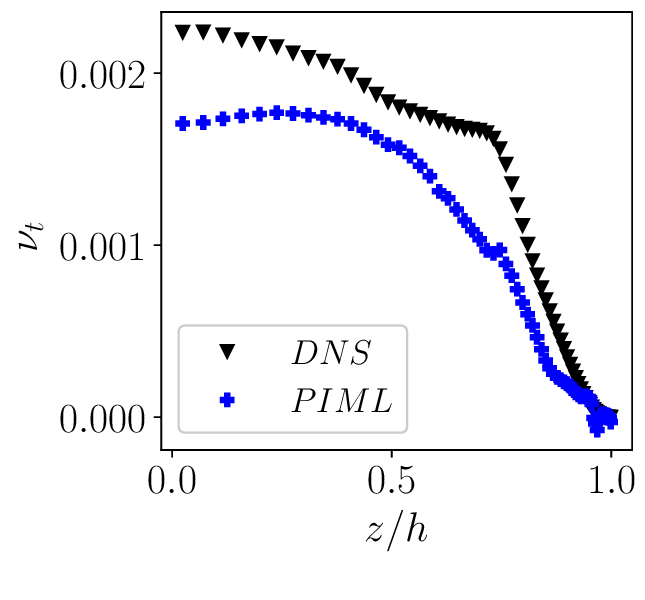}}%

\caption {Comparison of optimal turbulence viscosity profile
$\overline{u_y}$
obtained from DNS data and PIML simulation for Re=3500}
\label{fig:nu_t_Re3500} 
\end{figure}

\begin{figure}[h!]
\centering
\subfloat[component $\tau_{yy}$
Reynolds stress in the section
$y/h=0.25$
]{ \label{fig:tau_yy_Re3500_25}
\includegraphics[width=0.33\textwidth]{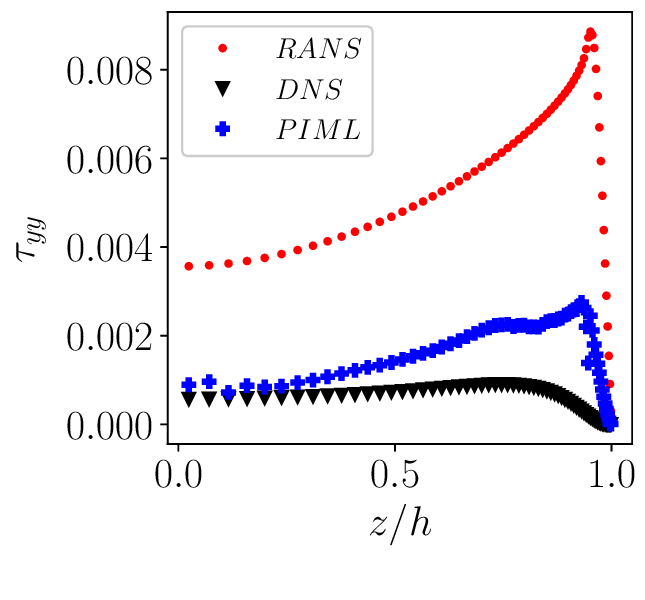}}
\subfloat[component $\tau_{yy}$
Reynolds stress in the section
$y/h=0.50$
]{ \label{fig:tau_yy_Re3500_50}
\includegraphics[width=0.33\textwidth]{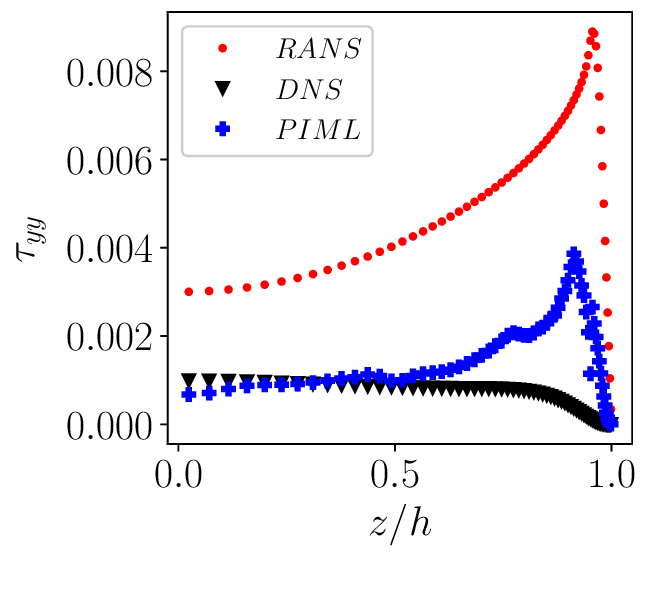}}%
\subfloat[component $\tau_{yy}$
Reynolds stress in the section
$y/h=0.75$
]{ \label{fig:tau_yy_Re3500_75}
\includegraphics[width=0.33\textwidth]{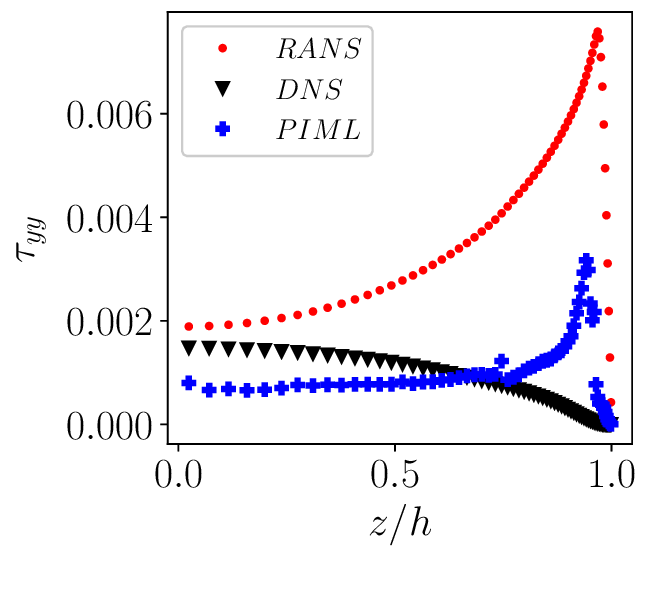}}%

\caption {Comparison of normal components in the secondary flow plane of Reynolds stress}
\label{fig:tau_yy_Re3500} 
\end{figure}

\begin{figure}[h!]
\centering
\subfloat[component $\tau_{zz}$
Reynolds stress in the section
$y/h=0.25$
]{ \label{fig:tau_zz_Re3500_25}
\includegraphics[width=0.33\textwidth]{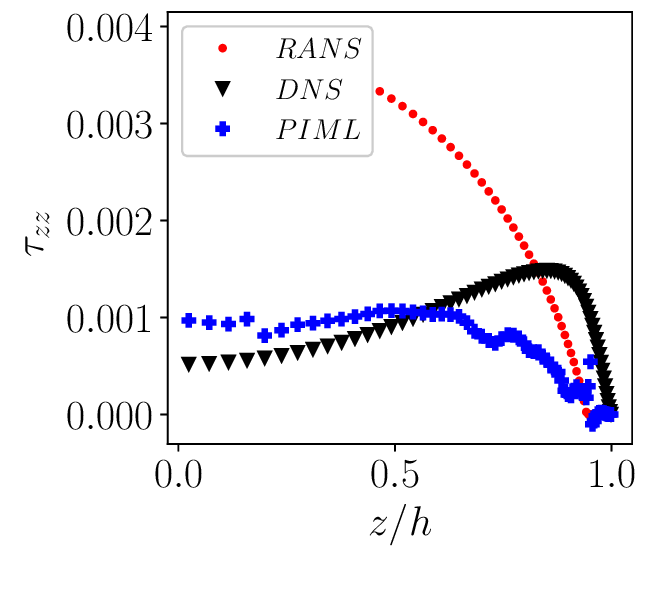}}
\subfloat[component $\tau_{zz}$
Reynolds stress in the section
$y/h=0.50$
]{ \label{fig:tau_zz_Re3500_50}
\includegraphics[width=0.33\textwidth]{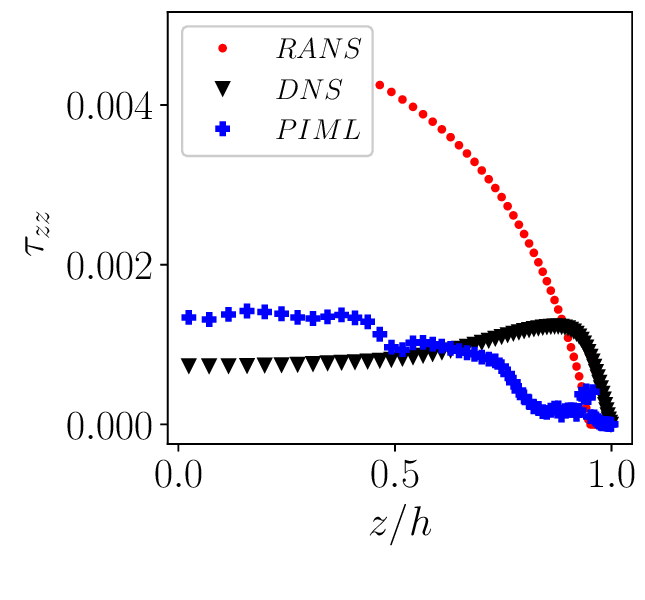}}%
\subfloat[component $\tau_{zz}$
Reynolds stress in the section
$y/h=0.75$
]{ \label{fig:tau_zz_Re3500_75}
\includegraphics[width=0.33\textwidth]{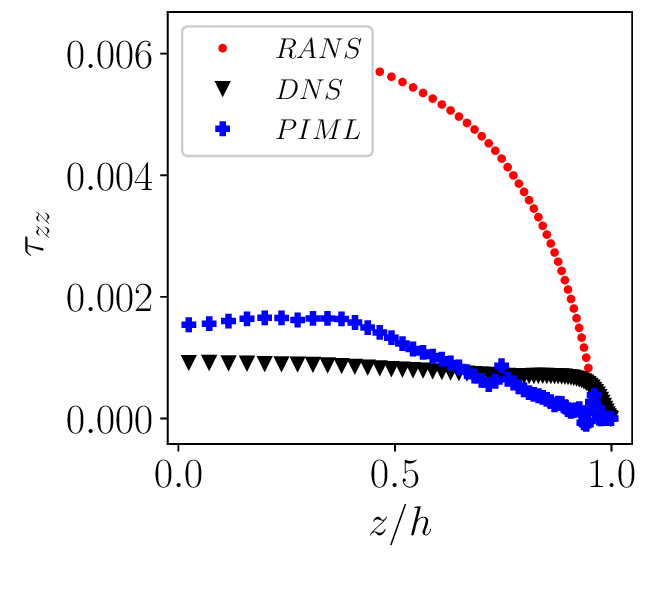}}%

\caption {Comparison of normal components in the secondary flow plane of Reynolds stress}
\label{fig:tau_zz_Re3500} 
\end{figure}

\begin{figure}[h!]
\centering
\subfloat[secondary flow obtained from RANS]{ \label{fig:sec_flow_Re3500_comparison:RANS}
\includegraphics[width=0.33\textwidth]{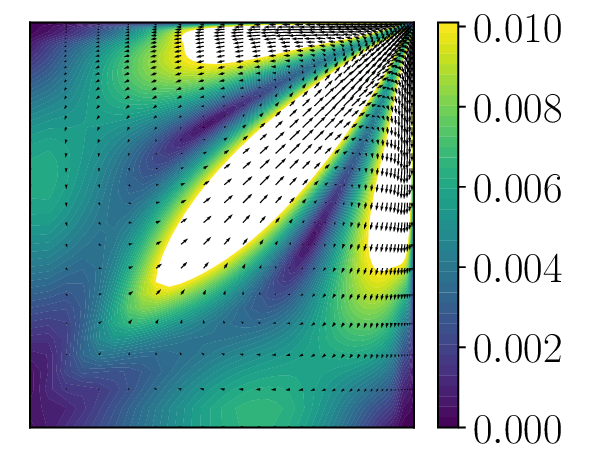}}%
\subfloat[secondary flow obtained from PIML]{ \label{fig:sec_flow_Re3500_comparison:PIML}
\includegraphics[width=0.33\textwidth]{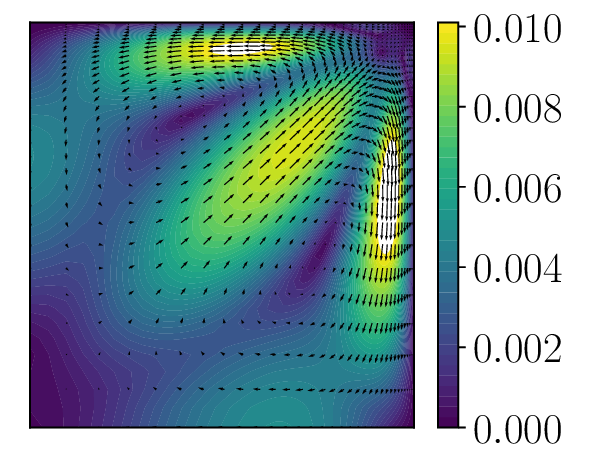}}%
\subfloat[secondary flow obtained from DNS]{ \label{fig:sec_flow_Re3500_comparison:DNS}
\includegraphics[width=0.33\textwidth]{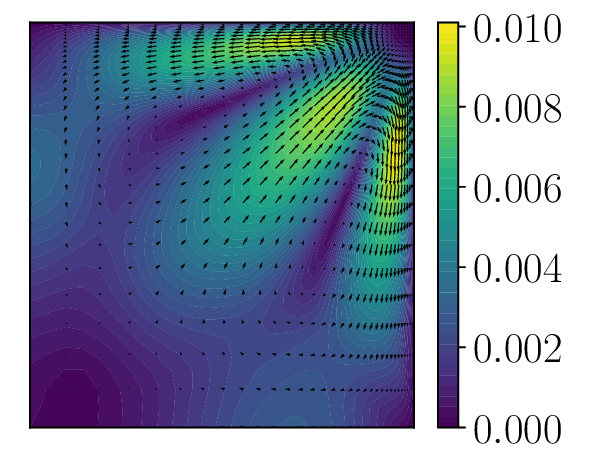}}
\caption{Comparison of secondary flow obtained from RANS, DNS and PIML simulations for
$Re=3500$
(The colored column numbers indicate the magnitude of the secondary velocity vector)}
\label{fig:sec_flow_Re3500_comparison} 
\end{figure}

\newpage
\pagebreak
\clearpage

\subsection{Available DNS Datasets and Turbase Converging-Diverging Channel}

A comprehensive search of available DNS datasets has been conducted, categorizing them based on whether the reported variables are time-averaged or provided as a time series. The physical similarity between the datasets and the target problem is also considered, excluding cases involving energy, state, or multiphase flow equations. Among the identified datasets, the flow in a converging-diverging channel was selected due to its relevance. The physics of blood flow in the target nozzle closely resembles the behavior in a converging-diverging channel, making it suitable for training a network to improve RANS simulations.

\begin{figure}[h!]
\centering
\includegraphics[width=0.8\textwidth]{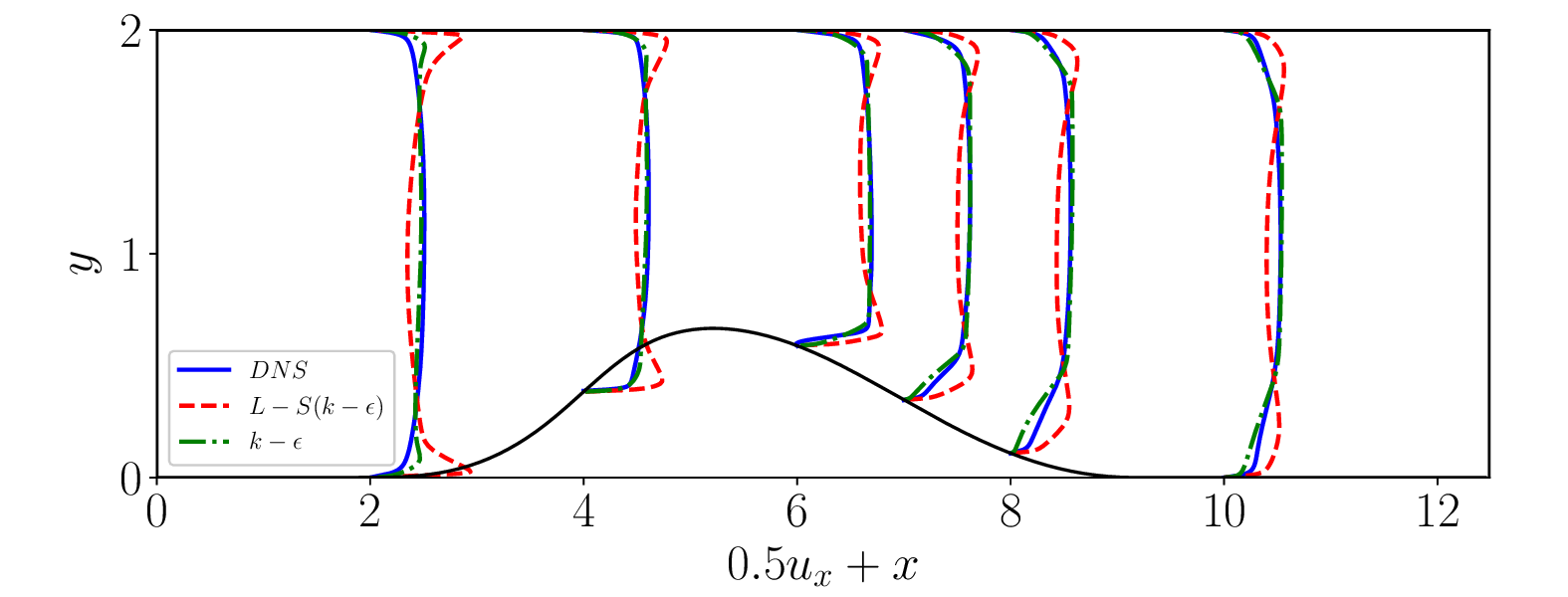}
\caption {Comparison of velocity profiles along the stream obtained from two RANS models and DNS data at different times}
\label{fig:Turbase_DNS_RANS_Velocity}
\end{figure}

As part of this work, DNS data from the European project WALLTURB was chosen as the reference dataset. To provide input features for the PIML network, RANS simulations must first be performed. The selected dataset contains data from two direct numerical simulations of flow within a convergent-divergent channel. These cases were designed using a geometry similar to that employed by the Lille Mechanics Laboratory (LML) in various experiments focused on turbulence analysis under pressure gradients.
In the figure \ref{fig:Turbase_DNS_RANS_Velocity} RANS and DNS velocity fields are shown for several sections simultaneously in terms of distance from the wall.

By taking similar approach as we took for the square-cross section channel flow, difference of mapped eigenvalues to the barycentric coordinate have been calculated and the variations over the domain are shown in below.

By applying the t-SNE dimensionality reduction technique, the feature set $G$ has been transformed into a two-dimensional plane. The points of this Cartesian plane are features. These points only specify the location of the features, and after determining the location, Hall's convex hull algorithm is used to find the smallest rectangle containing all the points. Since the image must be horizontal or vertical for the convolutional neural network architecture, a rotation is performed. Then, the features are averaged to convert the Cartesian coordinates to values per pixel of the image.
In order to use this method,a hyper-parameter called multiplexity, which determines the effect of local and non-local aspects of the data, must be determined. This parameter is, in a sense, a guess about the number of nearest neighbors of each point. 

\begin{figure}[h!]
\centering
\includegraphics[width=\textwidth]{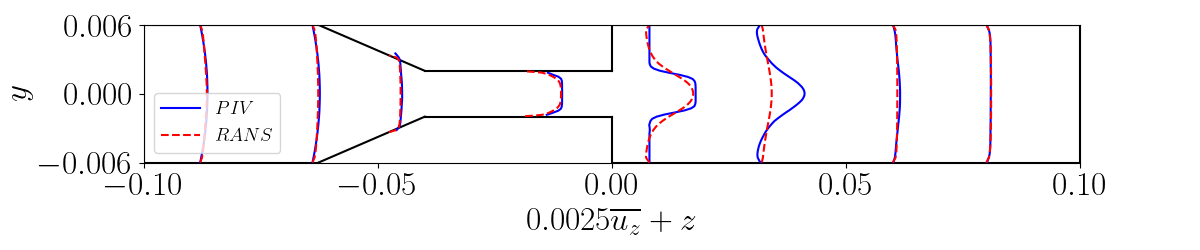}
\caption{$\overline{u_z}$ obtained from RANS and PIV}
\label{fig:FDA_PIV_RANS_velocity}
\end{figure}

In Fig. \ref{fig:FDA_PIV_RANS_velocity}, the velocity along the z-axis obtained from the RANS simulation is compared with PIV laboratory results at different cross-sections.  

Viscous shear stress magnitude in an incompressible Newtonian fluid is calculated according to the following equation:
\begin{equation}
|\sigma| = 2 \mu |S_{ij}S_{ij}|
\end{equation}

These figures illustrate the complexities of shear stress behavior in the studied device, reiterating the need for precision in RANS modeling to ensure the high accuracy of predicted blood damage factors.
\begin{figure}[h!]
\centering
\subfloat[]{ \label{fig:Turbase_Delta_xi_eta:xi}
\includegraphics[width=0.8\textwidth]{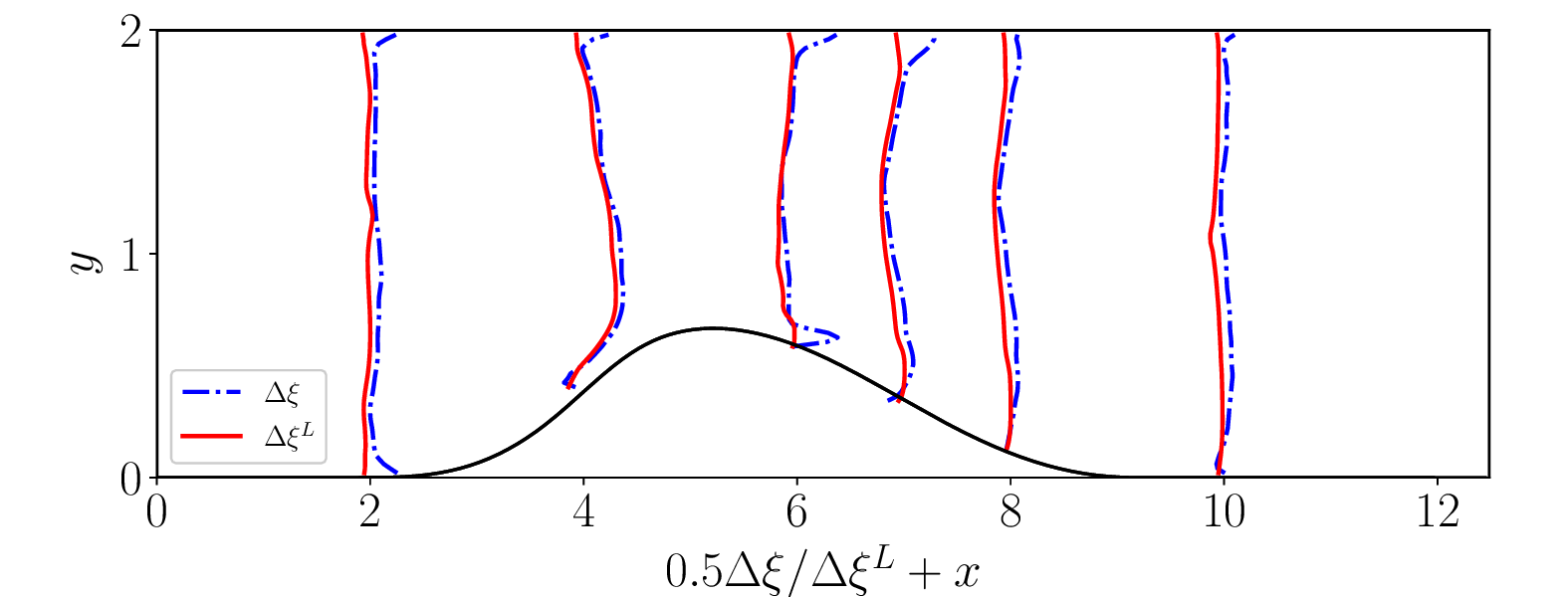}}
\hfill
\subfloat[]{ \label{fig:Turbase_Delta_xi_eta:eta}
\includegraphics[width=0.8\textwidth]{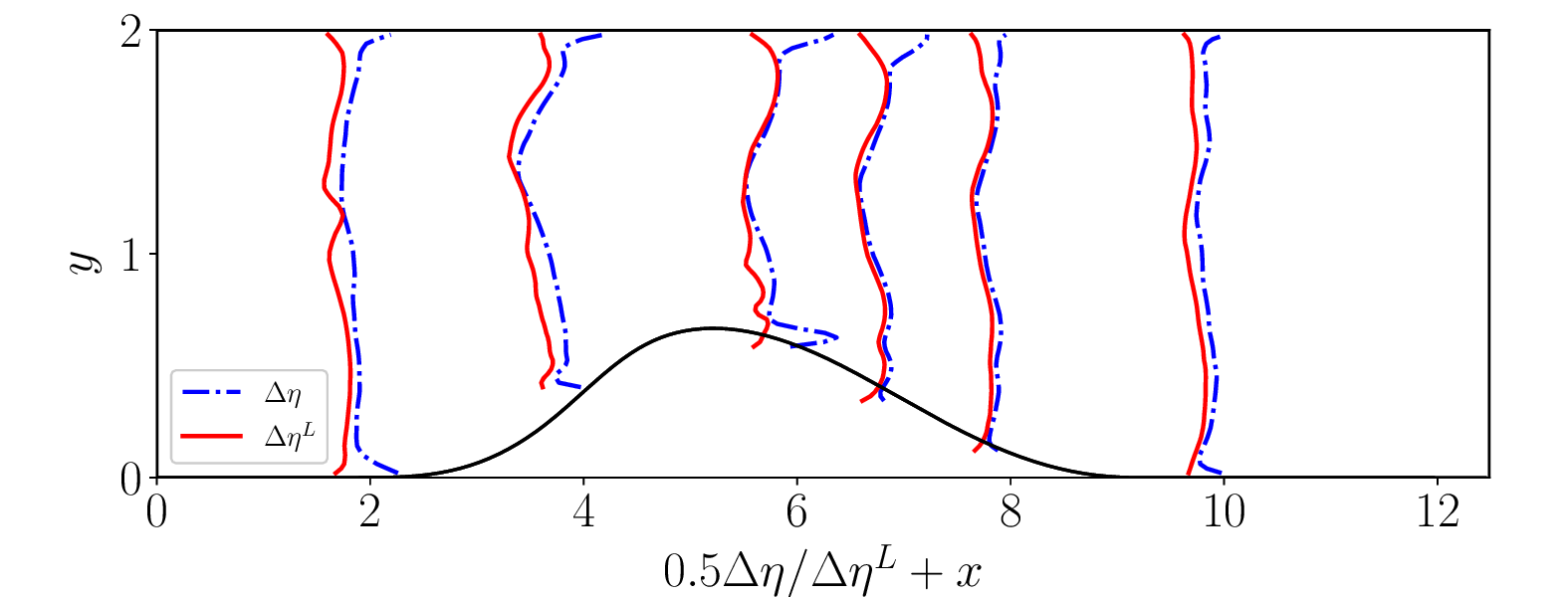}}

\caption {Comparison of eigenvalues of the deviatoric part of Reynolds stress mapped on barycentric coordinates}
\label{fig:Turbase_Delta_xi_eta} 
\end{figure}

\begin{figure}[h!]
\centering
\includegraphics[width=0.8\textwidth]{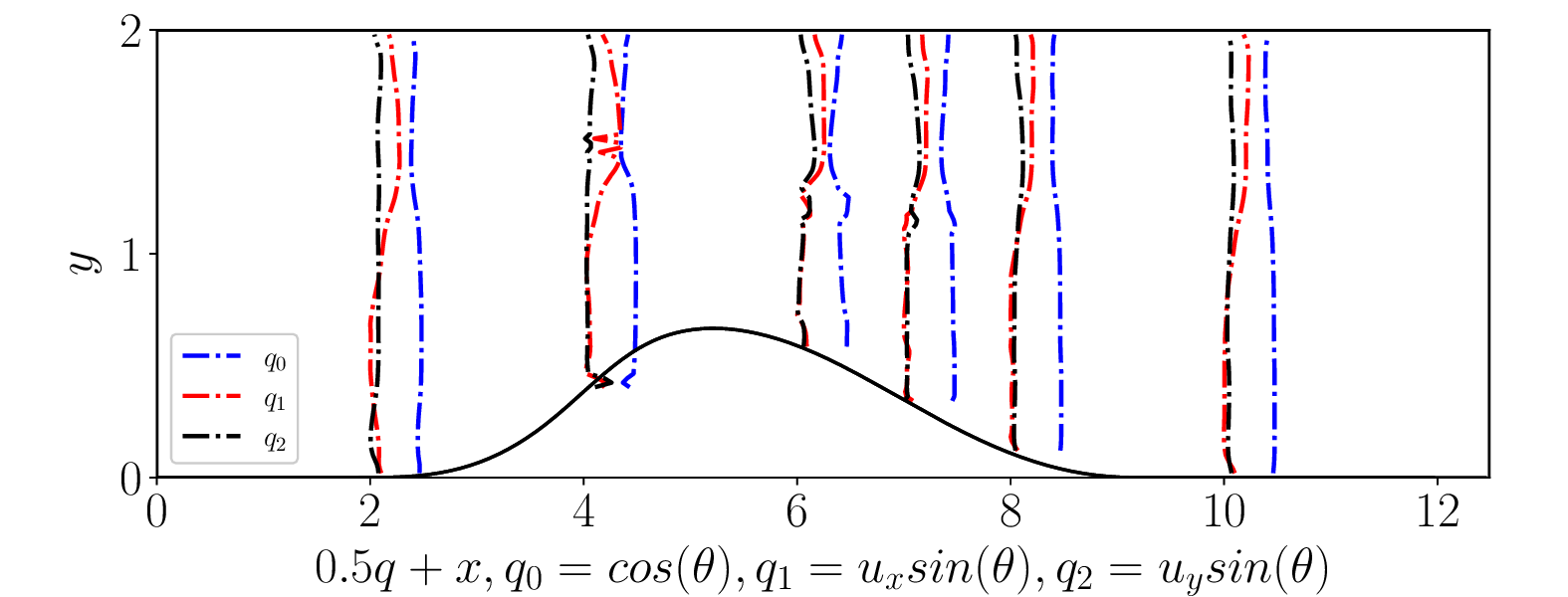}
\label{fig:Turbase_q}
\caption {The first three components of the proper period quaternion for the rotation of DNS eigenvectors to RANS}
\end{figure}

\begin{figure}[h!]
\centering
\includegraphics[width=0.9\textwidth]{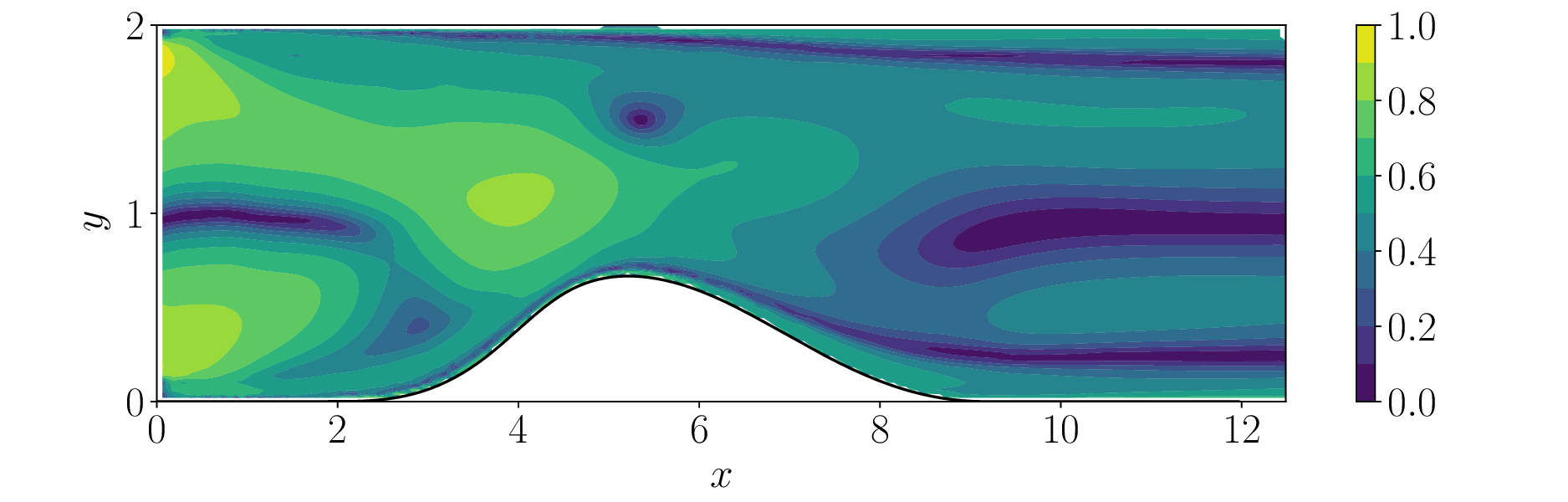}
\caption{$tr(\hat{\mathbf{S}}^2)$ obtained from RANS simulation}
\label{fig:Turbase_input_modified0}
\end{figure}

\begin{figure}[h!]
\centering
\includegraphics[width=0.9\textwidth]{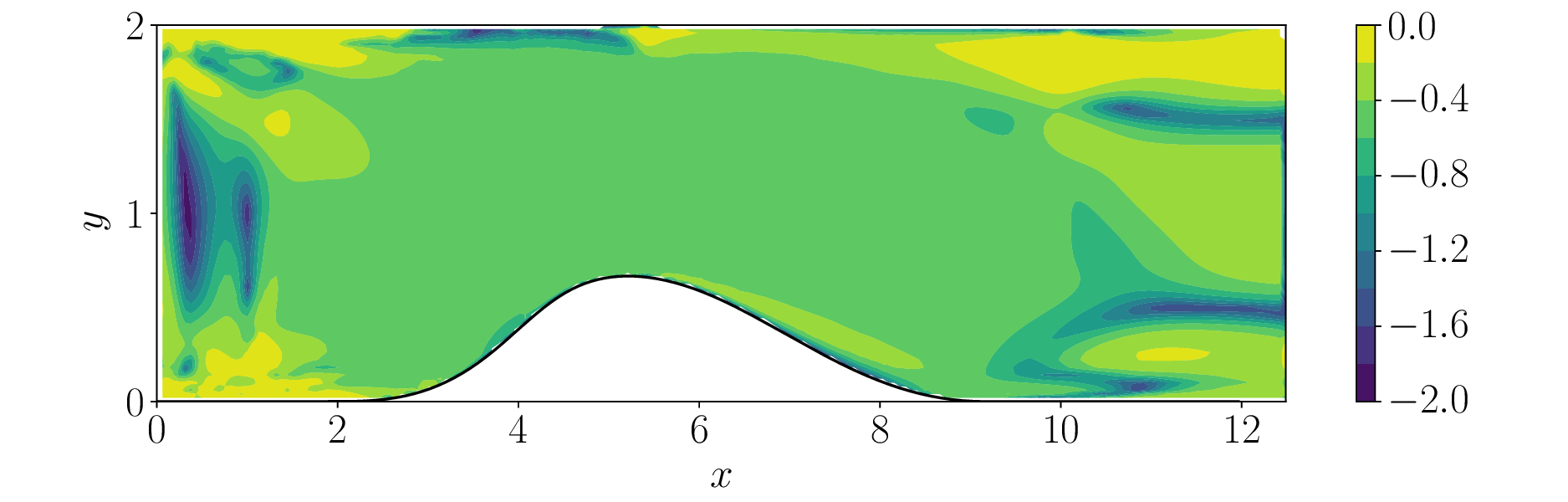}
\label{fig:Turbase_input_modified3}
\caption{$tr(\hat{\mathbf{A_p}}^2)$ obtained from RANS simulation}
\end{figure}

\begin{figure}[h!]
\centering
\subfloat[]{ \label{fig:Turbase_train_all_tsne:7}
\includegraphics[width=0.45\textwidth]{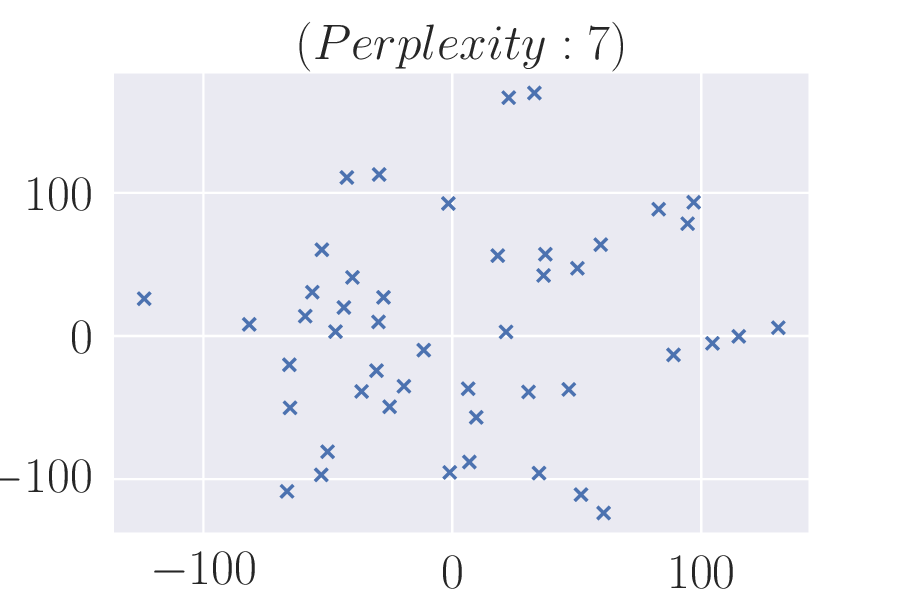}}
\hfill
\subfloat[]{ \label{fig:Turbase_train_all_tsne:12}
\includegraphics[width=0.45\textwidth]{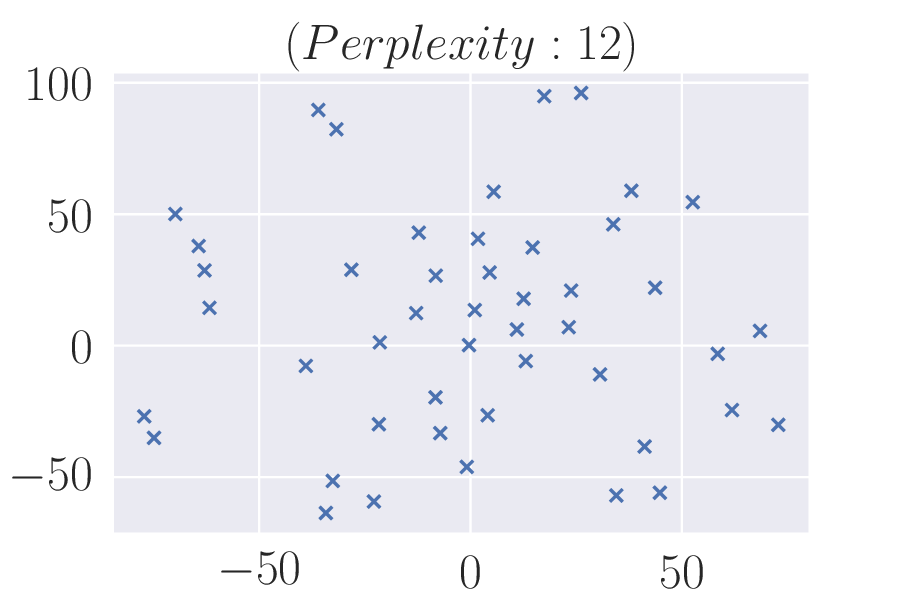}}

\caption{Conversion of feature set on two-dimensional screen with t-SNE dimension reduction method}
\label{fig:Turbase_train_all_tsne} 
\end{figure}

\begin{figure}[h!]
\centering
\subfloat[]{ \label{fig:Turbase_TSNE:7}
\includegraphics[width=0.45\textwidth]{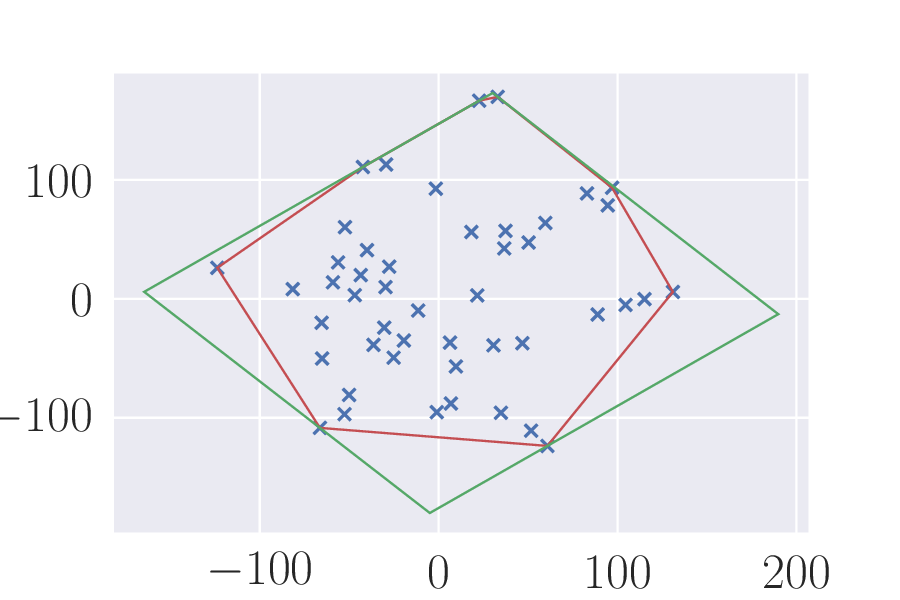}}
\hfill
\subfloat[]{ \label{fig:Turbase_TSNE:12}
\includegraphics[width=0.45\textwidth]{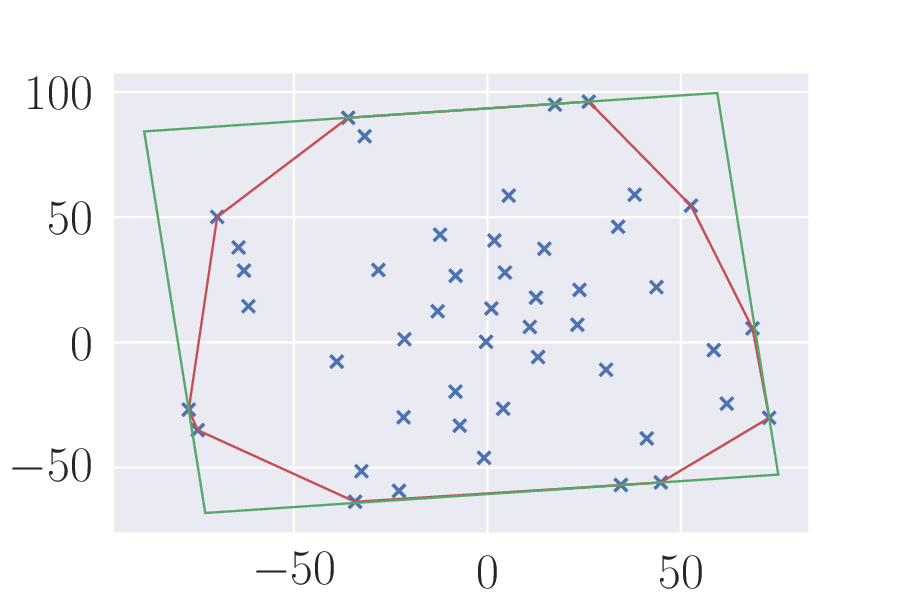}}

\caption {Applying Hall's convex hull algorithm to find the smallest rectangle containing all the features}
\label{fig:Turbase_TSNE} 
\end{figure}
\begin{figure}[h!]
\centering
\subfloat[]{ \label{fig:Turbase_Feature_Density_Matrix:7}
\includegraphics[width=0.45\textwidth]{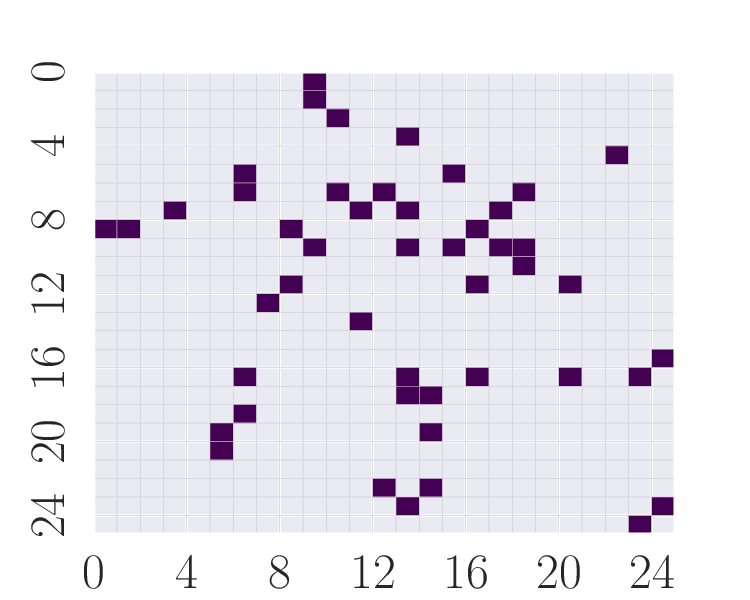}}
\hfill
\subfloat[]{ \label{fig:Turbase_Feature_Density_Matrix:12}
\includegraphics[width=0.45\textwidth]{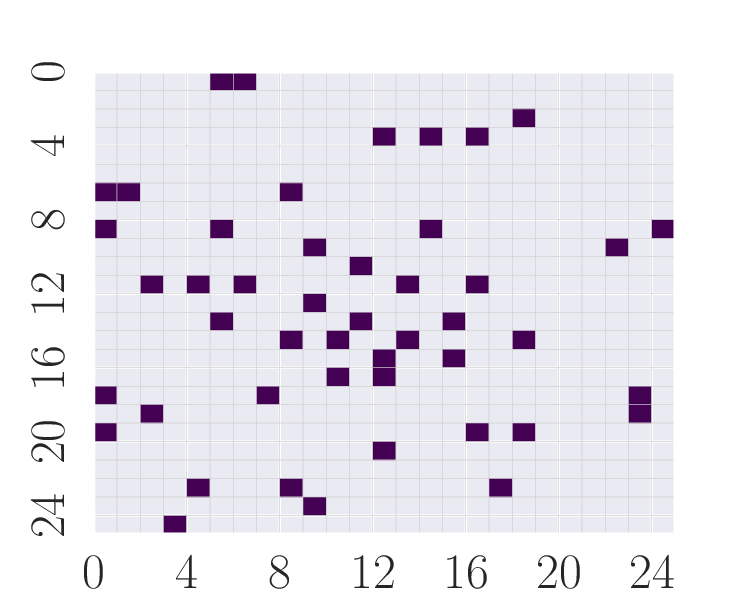}}
\caption{Feature density matrix}
\label{fig:Turbase_Feature_Density_Matrix} 
\end{figure}
\begin{figure}[h!]
\centering
\subfloat[]{ \label{fig:Turbase_DI_input_im:10000}
\includegraphics[width=0.33\textwidth]{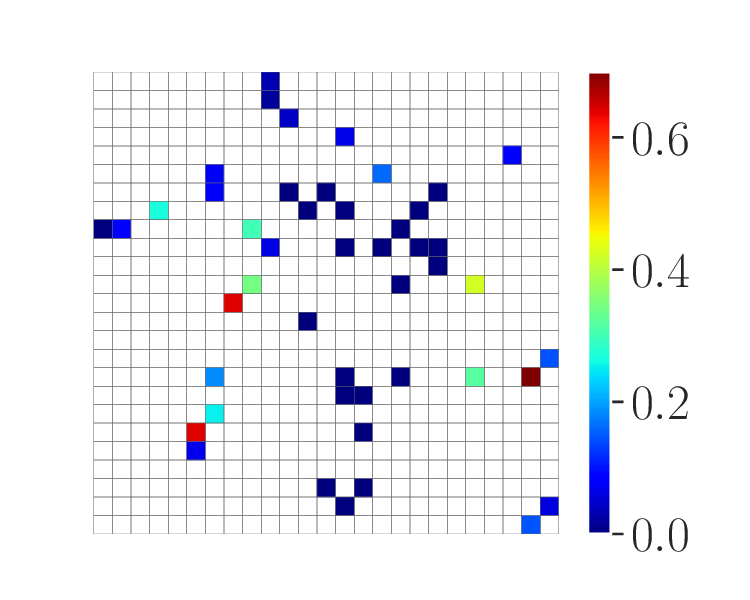}}
\subfloat[]{ \label{fig:Turbase_DI_input_im:20000}
\includegraphics[width=0.33\textwidth]{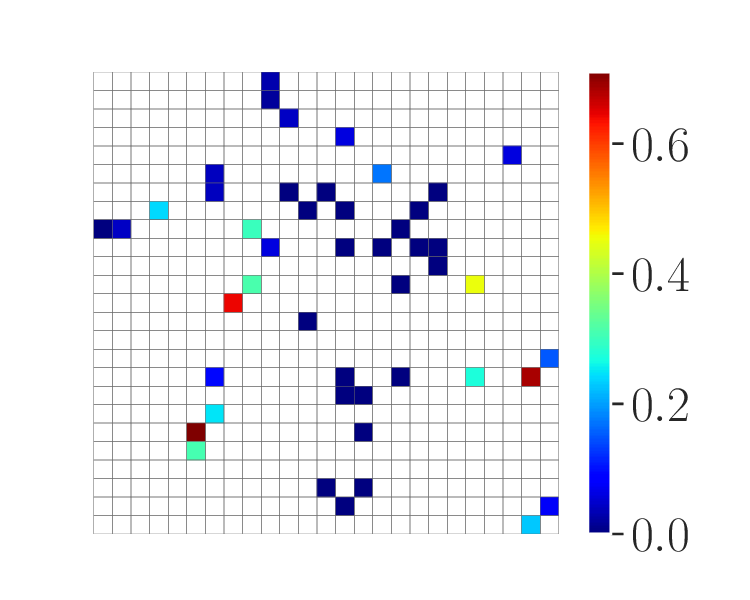}}
\subfloat[]{ \label{fig:Turbase_DI_input_im:30000}
\includegraphics[width=0.33\textwidth]{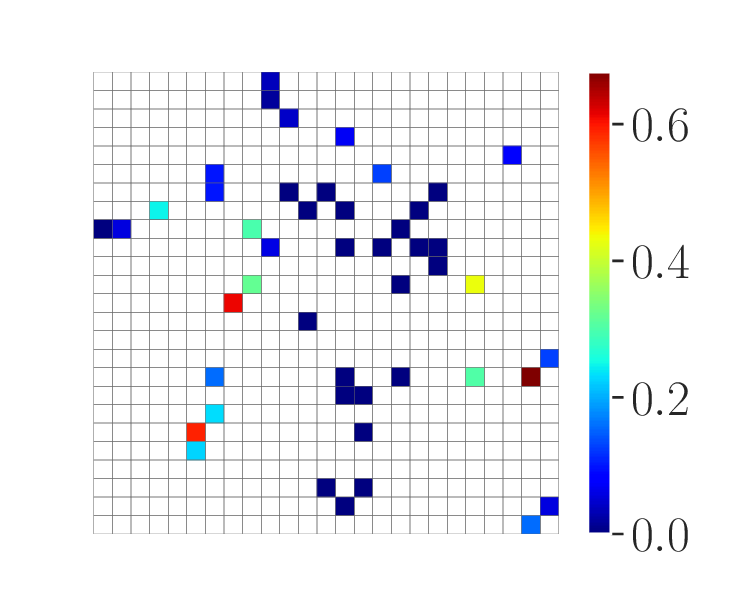}}
\hfill
\subfloat[]{ \label{fig:Turbase_DI_input_im:40000}
\includegraphics[width=0.33\textwidth]{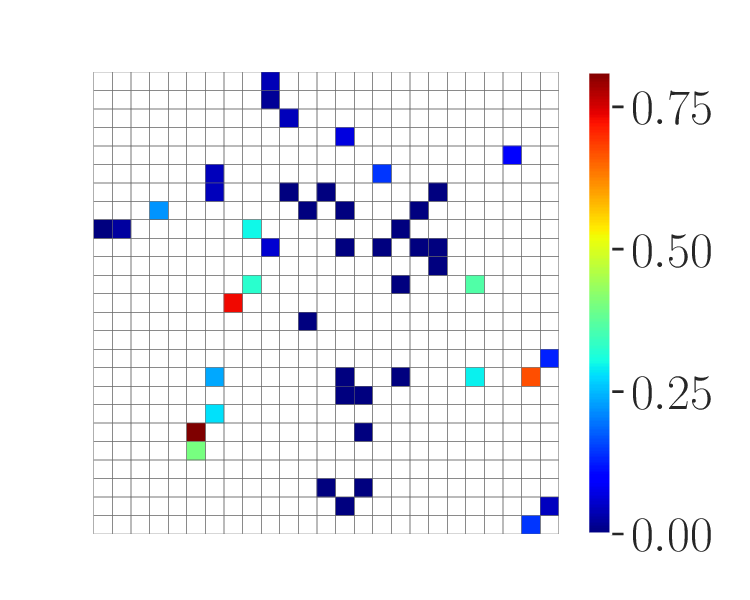}}
\subfloat[]{ \label{fig:Turbase_DI_input_im:50000}
\includegraphics[width=0.33\textwidth]{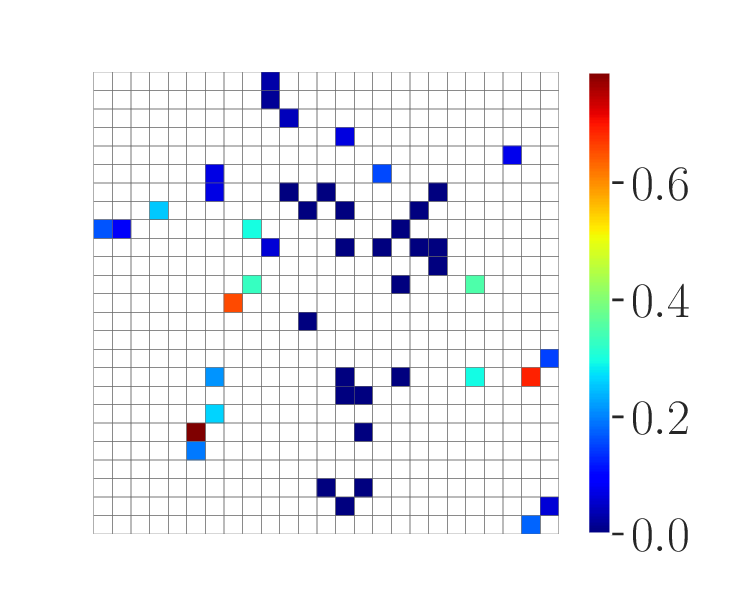}}
\subfloat[]{ \label{fig:Turbase_DI_input_im:60000}
\includegraphics[width=0.33\textwidth]{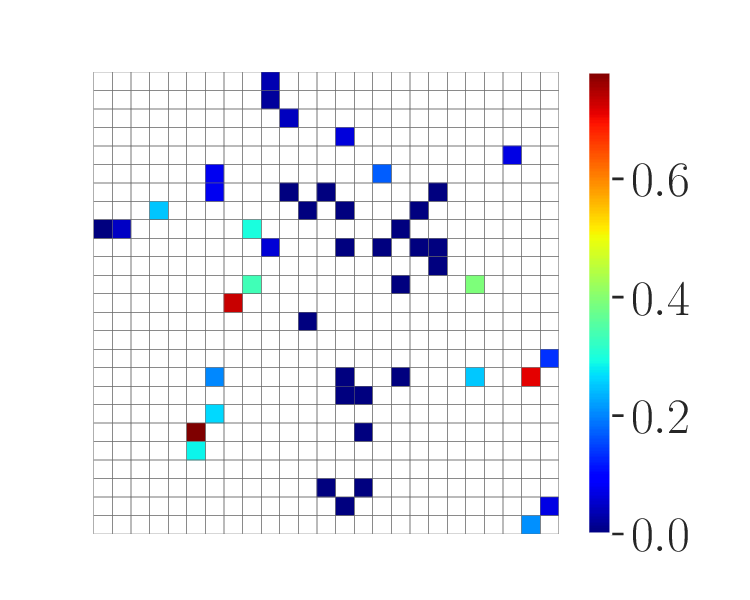}}

\caption{6 examples of images obtained with the DeepInsight method}
\label{fig:Turbase_DI_input_im} 
\end{figure}

\begin{figure}[h!]
\centering
\includegraphics[width=\textwidth]{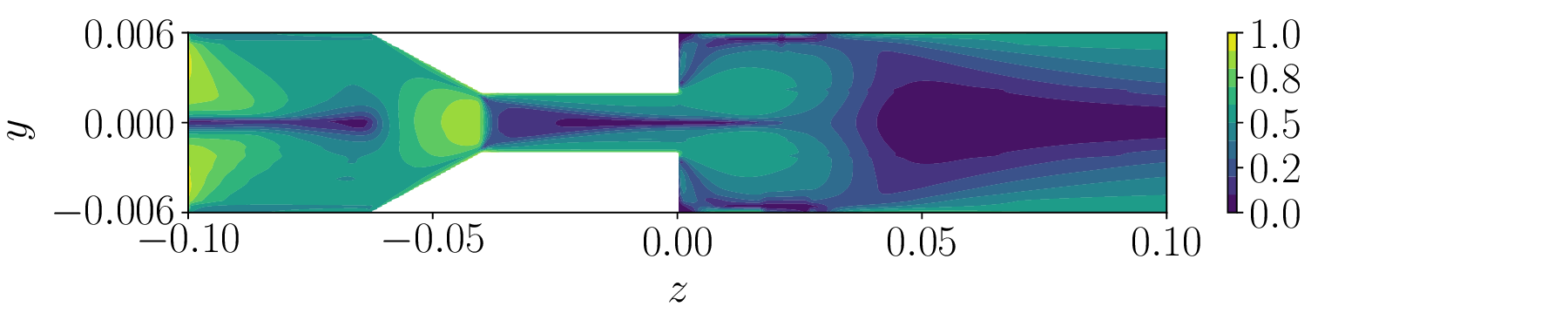}
\caption{$tr(\hat{\mathbf{S}}^2)$ obtained from RANS simulation}
\label{fig:FDA_input_i_0}
\end{figure}
\begin{figure}[h!]
\centering
\includegraphics[width=\textwidth]{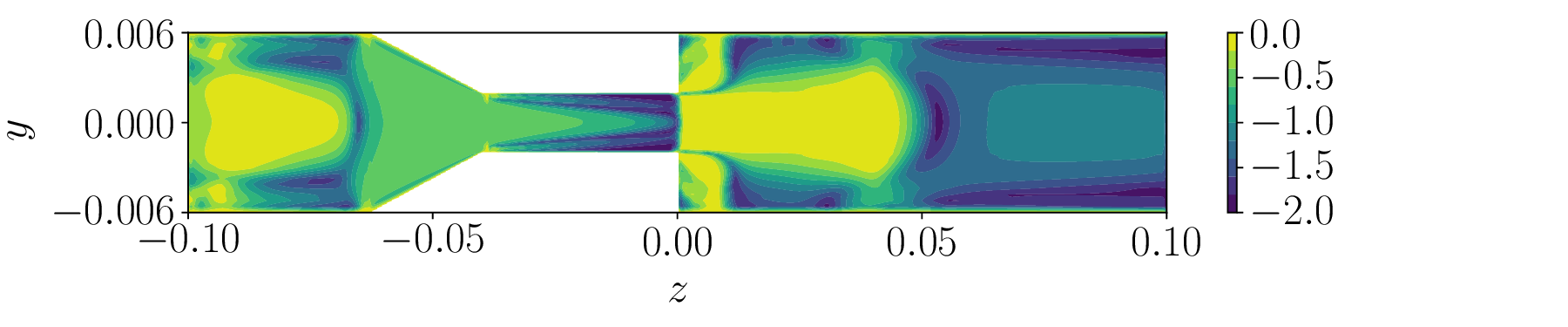}
\label{fig:FDA_input_i_3}
\caption{$tr(\hat{\mathbf{A_p}}^2)$ obtained from RANS simulation}
\end{figure}
\begin{figure}[h!]
\centering
\includegraphics[width=\textwidth]{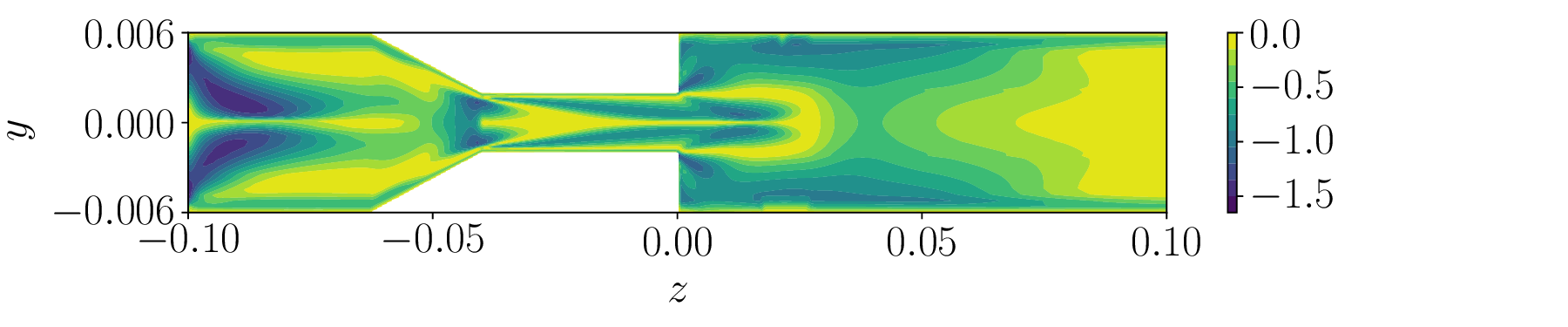}
\label{fig:FDA_input_i_4}
\caption{$tr(\hat{\mathbf{A_k}}^2)$ obtained from RANS simulation}
\end{figure}
\begin{figure}[h!]
\centering
\includegraphics[width=\textwidth]{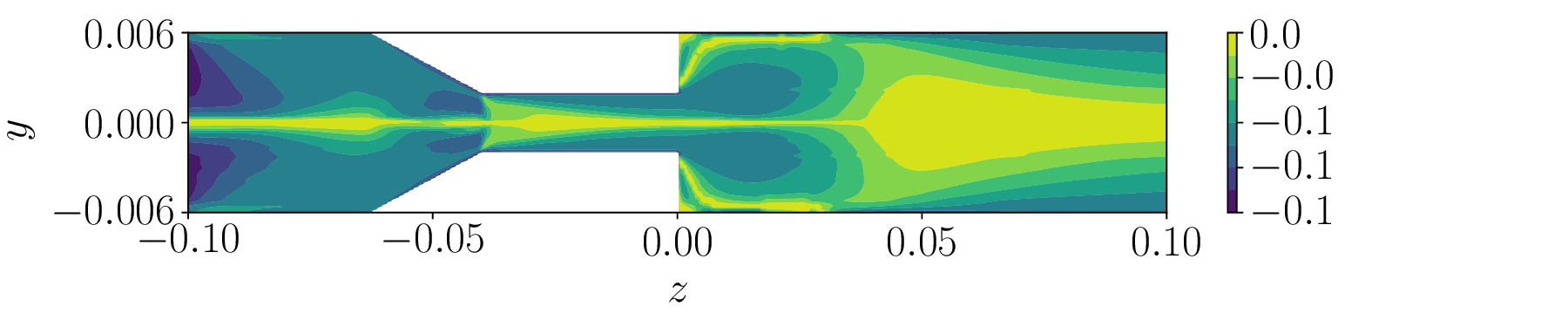}
\label{fig:FDA_input_i_6}
\caption{$tr(\hat{\mathbf{\Omega}}^2 \hat{\mathbf{S}}^2)$ obtained from RANS simulation}
\end{figure}
\begin{figure}[h!]
\centering
\includegraphics[width=\textwidth]{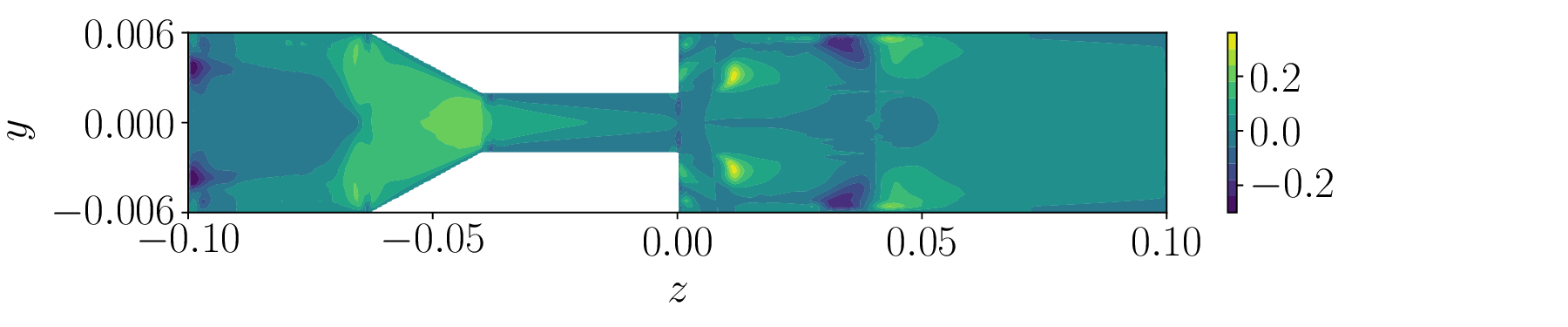}
\caption{$tr(\hat{\mathbf{A_p}}^2 \hat{\mathbf{S}})$ obtained from RANS simulation}
\label{fig:FDA_input_i_8}
\end{figure}


\begin{figure}[h!]
\centering
\includegraphics[width=\textwidth]{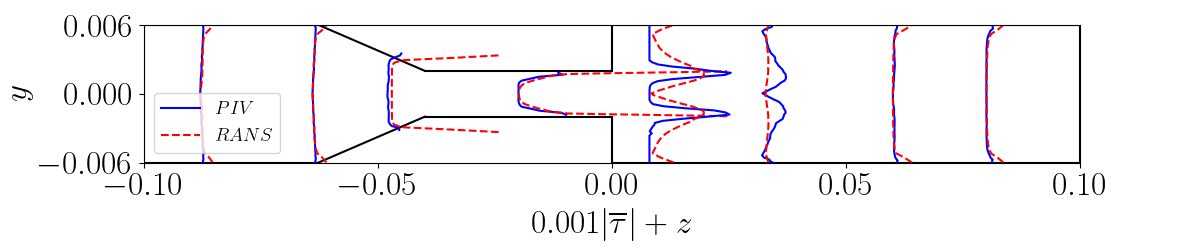}
\caption {Viscous shear stress obtained from RANS and PIV}
\label{fig:FDA_PIV_RANS_shear_stress}
\end{figure}
\begin{figure}[h!]
\centering
\includegraphics[width=\textwidth]{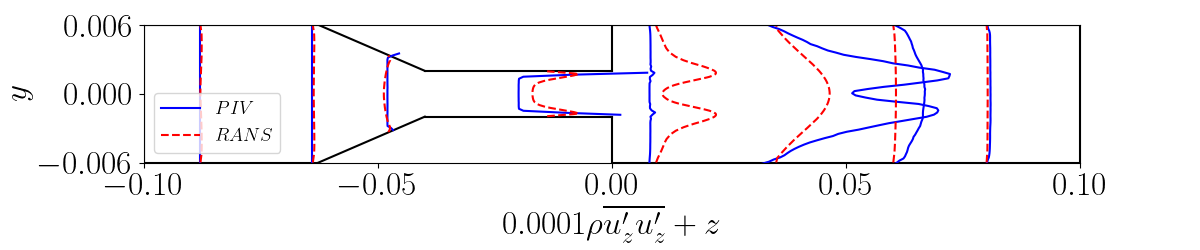}
\caption{Axial Reynolds stress obtained from RANS and PIV}
\label{fig:FDA_PIV_RANS_Reynolds_stress_zz}
\end{figure}
\begin{figure}[h!]
\centering
\includegraphics[width=\textwidth]{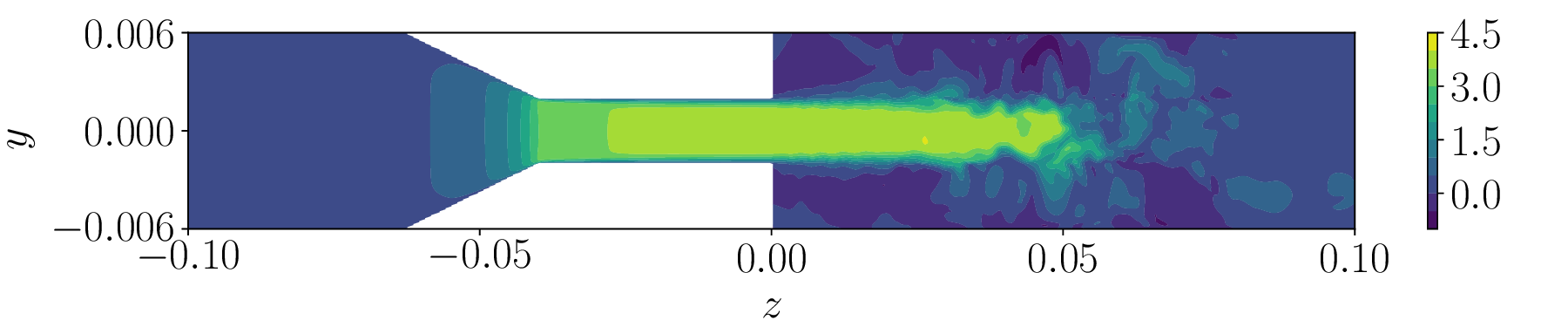}
\caption{ Velocity profile obtained from PIML after 100 iterations}
\label{fig:FDA_RANS_Velocity_BL3e4_13}
\end{figure}

\begin{figure}[h!]
\centering
\includegraphics[width=\textwidth]{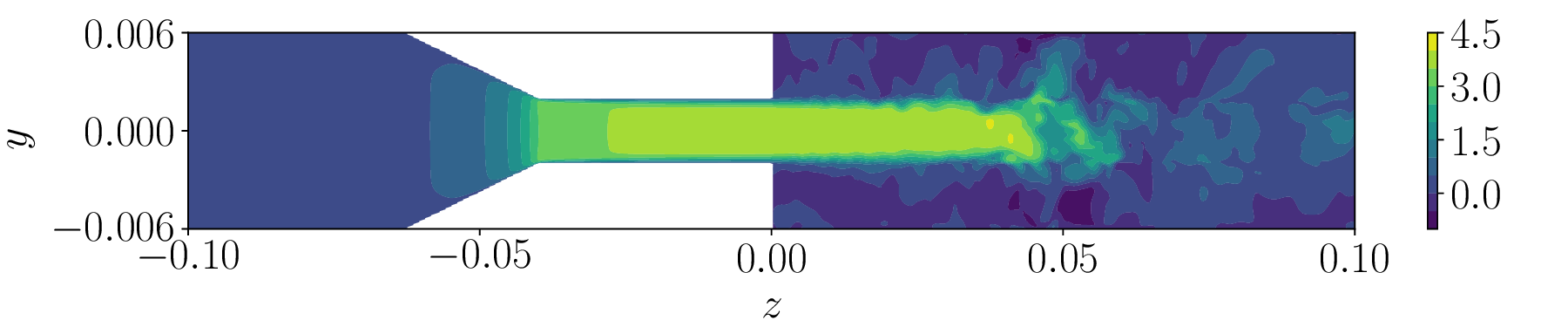}
\caption{ Velocity profile obtained from PIML after 150 iterations}
\label{fig:FDA_RANS_Velocity_BL3e4_10}
\end{figure}

\begin{figure}[h!]
\centering
\includegraphics[width=\textwidth]{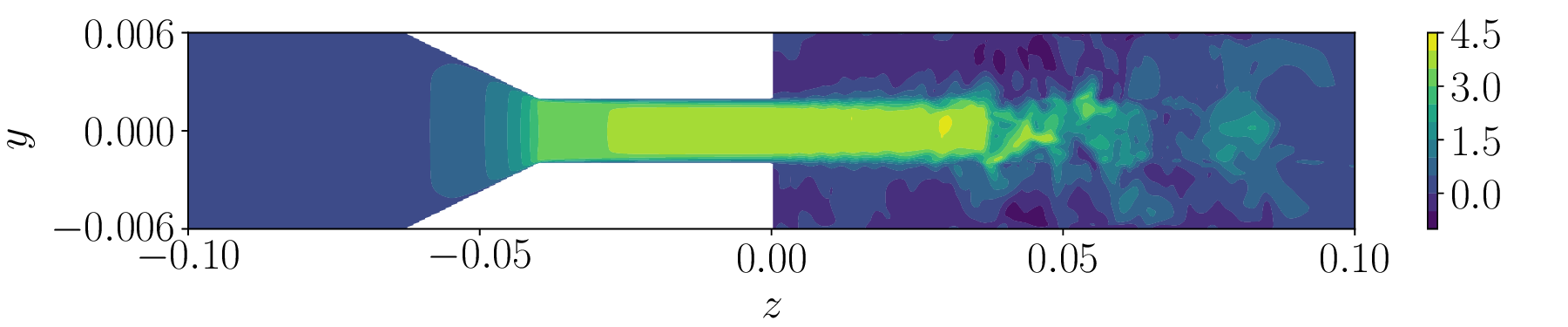}
\caption{ Velocity profile obtained from PIML after 200 iterations}
\label{fig:FDA_RANS_Velocity_BL3e4_8}
\end{figure}

\begin{figure}[h!]
\centering
\includegraphics[width=\textwidth]{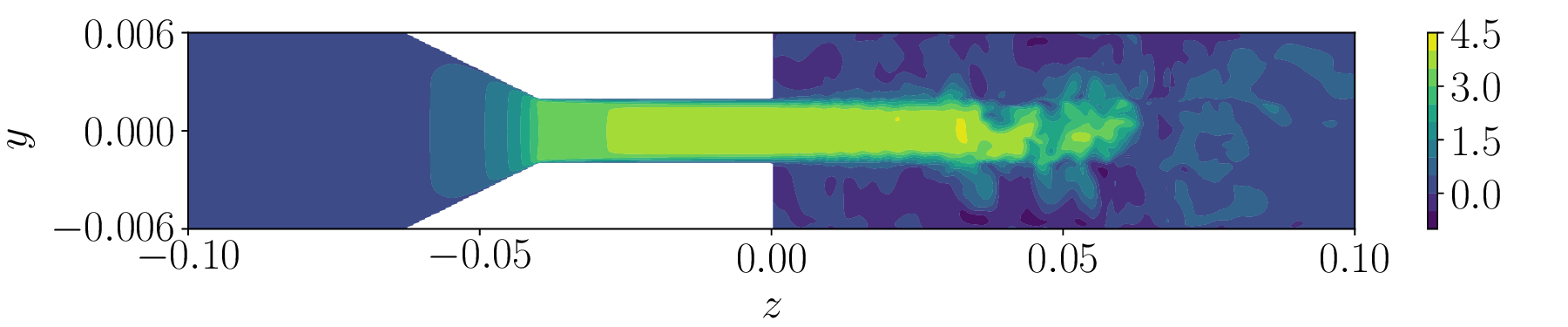}
\caption{ Velocity profile obtained from PIML after 400 iterations}
\label{fig:FDA_RANS_Velocity_BL3e4_6}
\end{figure}

\begin{figure}[h!]
\centering
\includegraphics[width=\textwidth]{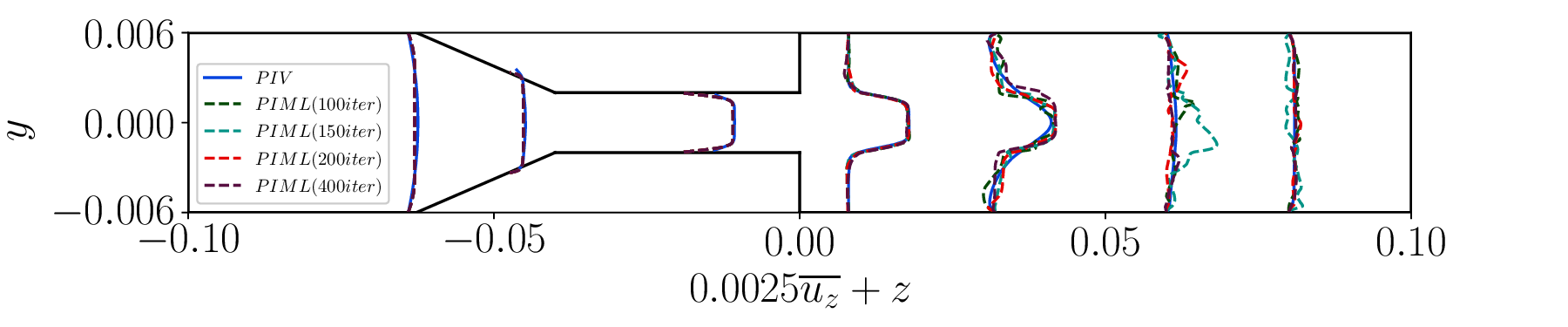}
\caption {Comparison of the speed profile obtained from PIML after different iterations and PIV data}
\label{fig:FDA_PIV_PIML_vel_prof_BL3e4_comparison}
\end{figure}

\begin{figure}[h!]
\centering
\includegraphics[width=\textwidth]{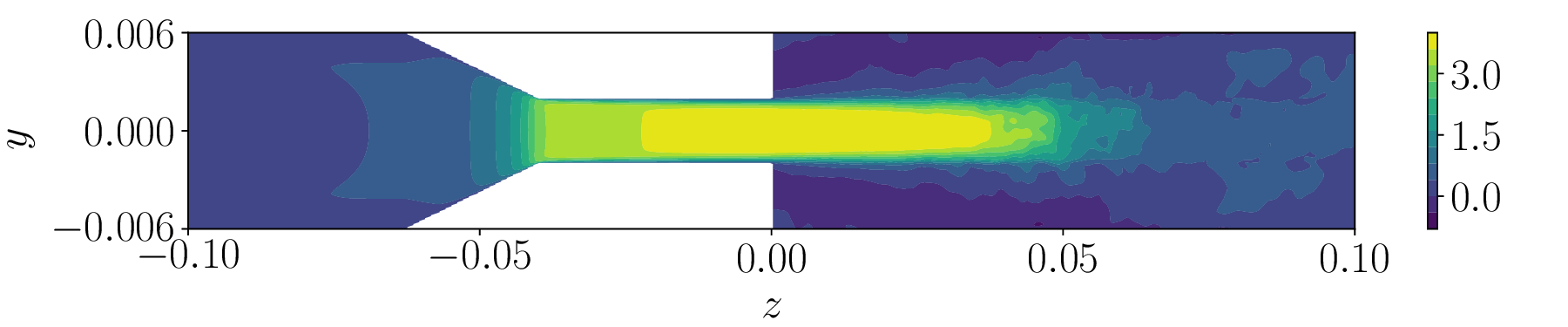}
\caption {Speed profile obtained from group averaging up to 5000 repetitions of different iterations of PIML solution}
\label{fig:FDA_RANS_Velocity_BL3e4_avg}
\end{figure}

\begin{figure}[h!]
\centering
\includegraphics[width=\textwidth]{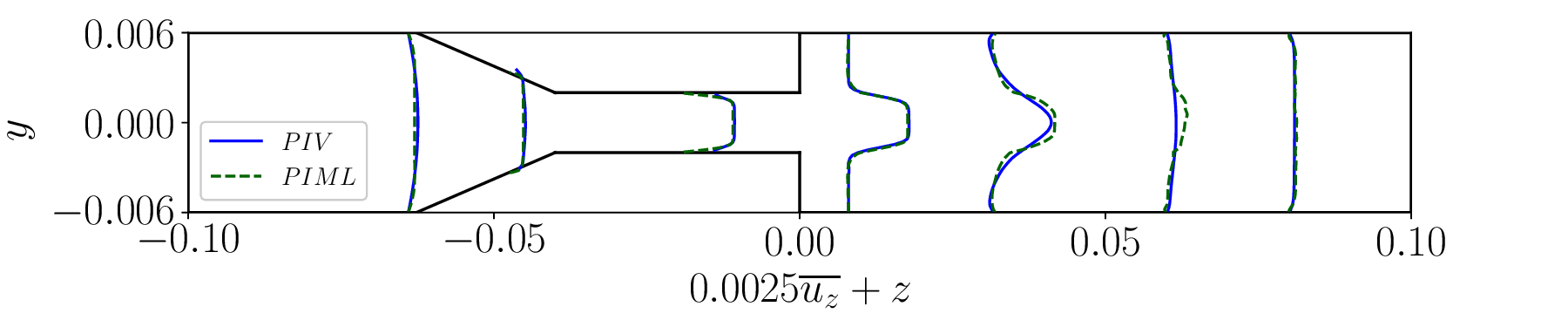}
\caption {Comparison of the speed profile obtained from PIML by applying group averaging up to 5000 repetitions and PIV}
\label{fig:FDA_PIV_PIML_vel_prof_BL3e4_avg}
\end{figure}

\pagebreak
\clearpage
\section{Conclusion}
In this paper we tried to improve RANS simulation by taking advantage of the high potential of data-oriented learning algorithms. For this purpose, a new solver based on the modified RANS equation with 3 correction variables was developed. The first correction variable was included in an implicit expression with the title of optimal turbulence viscosity. Two other corrective variables were entered into the flow equations directly from the artificial neural network.
To check the accuracy of the performance of the developed algorithm and solver, the flow in the channel with a square section was chosen. This choice was made due to the shortcomings of the common RANS models in simulating the secondary flows in the channel with a square section. By using DNS data at a lower Reynolds number of 2900 and using a multi-layer perceptron neural network, the invariant features obtained from RANS as input, and the angles and rotation vectors of the transformation of the eigenvectors of the Reynolds stress tensor RANS to DNS, the barycentric differences of the eigenvalues of the two stress matrices Reynolds RANS and DNS, as well as optimal turbulence viscosity were used as outputs of the multilayer perceptron network. By applying the trained network on a higher Reynolds number of 3500, the modified flow field was obtained and by comparing with the DNS simulation data, the good performance of the algorithm was observed. This observation can be promising for the application of this approach in reducing the cost of calculations by reducing the number of elements of the solution network.

Considering that in DNS simulation, the distance of the first cell from the wall is determined depending on the flow conditions such as the Reynolds number, with the present algorithm, the flow in the desired geometry can be simulated at a lower Reynolds number with fewer elements, and from its results to train the network Artificial neural and combined with the developed algorithm used in this thesis.

Next, the comprehensiveness and extrapolation capability of the learned network with DNS data of problems with different geometries from the target problem was investigated. It was investigated by considering 3 different geometries, each with similar flow physics to the 3 sections of the FDA standard nozzle. For all three problems, by performing RANS simulation, the feature vector was converted into a feature image with the DeepInsight algorithm. By examining the image of the features obtained for different points of the solution network, it was determined that the network outputs are sensitive to the input features. In fact, the features
  $\hat{\mathbf{\Omega}}^2\hat{\mathbf{S}}^2$
  And
$\hat{\mathbf{\Omega}}\hat{\mathbf{A_p}}\hat{\mathbf{S}}^2$
   And
  $\hat{\mathbf{\Omega}}\hat{\mathbf{A_k}}\hat{\mathbf{A_p}}\hat{\mathbf{S}}$
    The main representative of changes and variables with high sensitivity were identified in three basic learning problems. Also, by choosing the DeepInsight algorithm, it became possible to use the convolutional neural network.
Finally, the trained network was implemented on the US Food and Drug Administration standard nozzle problem, and by performing RANS simulation and forming input feature vectors, modified flow fields (determining viscous stresses) and Reynolds stress were obtained. The instant communication of the solver in each iteration with the artificial neural network to update the correction variables caused a quasi-unstable behavior in the final solution. By averaging the results of different iterations, we reached the flow field with a much smaller difference than the RANS simulation with the experimental results. The modified stress and flow field can be used to solve the hemolysis criterion transfer equation with the Eulerian approach and using the power model.

In order to follow up the current research, the following suggestions are made.
\begin{itemize}
\item {The use of the developed algorithm in reducing the computational costs of DNS simulation should be investigated more closely. In fact, by specifying the flow conditions (e.g. Reynolds number) of the target, DNS simulations can be performed at lower Reynolds numbers and the performance of the algorithm in predicting and reducing the computational cost due to increasing the size of the solution grid elements can be checked.} \\
\item {Also, by simulating DNS on benchmark issues that are not listed in the \ref{tab:DNS_Database} table, it is possible to generate new benchmark data and learn a new neural network with the ability to extrapolate to more diverse issues.}\\
\item {In hemolysis modeling, other models can be used instead of the power model. Also, the effect of the Lagrangian approach instead of the Eulerian approach in simulating the hemolysis criterion with the modified flow field should be investigated.
Loss-based approaches can also be used in the simulation of the hemolysis criterion in the turbulence flow regime.
}
\end{itemize}

\clearpage

\newpage



\end{document}